\documentclass[a4paper,fleqn,usenatbib]{mnras}

\usepackage{newtxtext,newtxmath}
\usepackage[T1]{fontenc}
\usepackage{ae,aecompl}

\usepackage{graphicx}	% Including figure files
\usepackage{amsmath}	% Advanced maths commands
\usepackage{amssymb}	% Extra maths symbols

\usepackage[caption=false]{subfig}
\graphicspath{{figures/}}

\usepackage{etoolbox}
\makeatletter
\patchcmd\@combinedblfloats{\box\@outputbox}{\unvbox\@outputbox}{}{\errmessage{\noexpand patch failed}}
\makeatother

\title[GW Bursts from Pulsars]{Gravitational Burst Radiation from Pulsars in the Galactic centre and stellar clusters}

\author[T. Kimpson et al.]{
	Tom Kimpson,$^{1}$\thanks{E-mail: tom.kimpson.16@ucl.ac.uk}
	Kinwah Wu ,$^{1}$
	and Silvia Zane $^{1}$
	\\
	$^{1}$ Mullard Space Science Laboratory, University College London. Holmbury St. Mary, Dorking, Surrey, RH5 6NT, UK
}

\date{Accepted XXX. Received YYY; in original form ZZZ}

\pubyear{2020}

\begin{document}
\label{firstpage}
\pagerange{\pageref{firstpage}--\pageref{lastpage}}
\maketitle

\begin{abstract}
Pulsars (PSRs) orbiting intermediate or supermassive black holes at the centre of galaxies and globular clusters are known as Extreme Mass Ratio Binaries (EMRBs) and have been identified as precision probes of strong-field GR. For appropriate orbital parameters, some of these systems may also emit gravitational radiation in a `burst-like' pattern. The observation of this burst radiation in conjunction with the electromagnetic radio timing signal would allow for multimessenger astronomy in strong-field gravitational regimes. In this work we investigate gravitational radiation from these PSR-EMRBs, calculating the waveforms and SNRs and explore the influence of this GW on the pulsar radio signal. We find that for typical PSR-EMRBs, gravitational burst radiation should be detectable from both the Galactic centre and the centre of stellar clusters, and that this radiation will not meaningfully affect the pulsar timing signal, allowing PSR-EMRB to remain `clean' test-beds of strong-field GR.
\end{abstract}


%%%%%%%%%%%%%%%%%%%%%%%%%%%%%%%%%%%%%%%%%%%%%%%%%%

%%%%%%%%%%%%%%%%% BODY OF PAPER %%%%%%%%%%%%%%%%%%

\section{Introduction}
The Galactic centre and the core of stellar clusters are known to be regions of exceptionally high stellar density; within the central parsec of the Milky Way there are expected to be up to $10^8$ stars, whilst in the centre of globular clusters the densities can be as high as $10^6$ per cubic parsec. As a  consequence, heavy bodies are expected to sink to the centre of theses clusters, due to the effects of dynamical friction and mass segregation. Moreover, frequent two-body interactions mean that these regions are known to preferentially form pulsars with millisecond periods (MSPs). Indeed, $\sim 90 \%$ of the observed pulsar population of dense globular clusters have periods $< $ 20 ms \citep{Camilo2005,Ransom2008}. Furthermore, the incidence of pulsars in dense globular clusters is observed to be greater by a factor of $10^3$ per unit mass compared to the more sparse Galactic disk \citep{Freire2013}. No MSPs have currently been detected in the Galactic centre due to the combination of interstellar scattering and the insufficient sensitivity of current radio telescopes. However, it is predicted that within the central parsec of Sgr A* there are up to $10^3 - 10^4$ MSPs \citep{Wharton2012, Rajwade2017} and the increased sensitivity of next generation radio facilities - such as the Square Kilometer Array \citep[SKA,][]{Keane2015} or the Five-hundred-meter Aperture Spherical Telescope \citep[FAST,][]{Nan2011} - will make this population accessible. \newline 

\noindent In addition to a numerous MSP population, it is likely that the nuclei of clusters host massive black holes (BHs). At the Galactic centre there is thought to exist a supermassive BH with mass $\sim 4 \times 10^6 M_{\odot}$ \citep{Gillessen2009, Boehle2016}. In addition, the ``$M$-$\sigma$ relation'' \citep{Ferrarese2000,Graham2019, Greene2019} - extrapolated to the low-mass range - suggests that globular clusters could host central BHs of intermediate mass (IMBH, $10^3 - 10^5 M_{\odot}$). Whilst the existence of IMBH is not as firmly established as supermassive BH, consilient strands from stellar kinematics \citep[e.g. ][]{Gebhardt2002, VDM2010, Feldmeier2013}, radiative accretion signatures \citep[e.g. ][]{Ulvestad2007} and pulsar dynamics \citep{K2017NAT} all suggest that globular clusters could host IMBH. For a complete review and discussion on the current observational evidence for IMBH, see \citet{Mezcua2017}. In addition to globular clusters and the Galactic centre, the heart of dwarf elliptical galaxies are also thought to contain intermediate or supermassive BHs - for example M32 has a BH of mass $\sim 3.4 \times 10^6 M_{\odot}$ \citep{Marel1997} \newline 

\noindent Systems in which a massive BH is orbited by a MSP with a short orbital period (e.g. $P \sim 0.1$ years for the  Galactic Centre) are known as  Extreme Mass Ratio Binaries (EMRBs). MSPs are unique astronomical probes due to their long-term  gyroscopic stability \citep[][]{Verbiest2009, Manchester2018} and the high-precision  measurements that can be made with radio timing of pulsars \citep[e.g.][]{Liu2011, Desvignes2016,Lazarus2016, Liu2018}. This natural apparatus of a stable, accurate clock in a extreme gravitational environment has been identified as an ideal precision probe of general relativity (GR) in the strong-field, non-linear regime \citep[see e.g. ][]{Kramer2004, Wang2008, Wang2009, Liu2012,Remmen2013, Nampalliwar2013, Singh2014,Kramer2016, Saxton2016, Li2018,Kimpson2019a, Kimpson2019b}. The key BH parameters  - the mass, spin, and quadrupole moment ($M, S, Q$ respectively) - are expected to be measured to a precision of $ \delta M \sim 10^{-5}, \delta S \sim 10^{-3}, \delta Q \sim 10^{4}$ \citep{Liu2012,Psaltis2016}. These measurements are orders of magnitude better than any other current method (e.g. VLTI observations of S2 stars, VLBI observations with the Event Horizon Telescope). With these parameters measured to sufficient precision it is then possible to explore fundamental questions of GR and the nature of BHs. For example, the Cosmic Censorship conjecture \citep{Penrose1969} imposes the constraint,
\begin{eqnarray}
a \equiv \frac{cS}{GM^2} \leq 1 \ .
\end{eqnarray}
Additionally, via the No Hair Theorem \citep{Hansen1974} all higher order multipole moments of a BH are expressible in terms of the lower-order moments $M$ and $S$. In this case,
\begin{eqnarray}
q \equiv \frac{c^4Q}{G^2M^3} = -a^2
\end{eqnarray}
Pulsar timing measurements of a MSP-EMRB system would allow the veracity of the cosmic Censorship Conjecture and No Hair theorem to be tested to better than 1 \% precision \citep[see e.g.][]{Kramer2004, Liu2012, Liu2014, Wex1999,Eatough2015}. MSP-EMRB are also exceptionally important from the perspective of astrophysics and would establish whether the central dark nuclei of star clusters are indeed astrophysical BHs as described by the Kerr solution, or else something more exotic such as a boson star \citep{Kleihaus2012} or a BH described by some deviation from the Kerr solution  \citep[`bumpy' black holes][]{Yagi2016}. The precision measurement of the BH mass in the centre of nearby galaxies would also enforce strong constrains on the low end of the $M-\sigma$ relation \citep{Ferrarese2000}. If an MSP could be detected orbiting an IMBH, this would also naturally settle the debate on the existence of intermediate mass, astrophysical black holes \citep{Singh2014}. Indeed, given the precision with which the BH mass can be measured using a MSP-EMRB \citep[$\sim 10^{-5}$, ][]{Liu2012} only one MSP-EMRB system needs to be detected in order to shed light on these questions. In addition, pulsar timing of EMRBs can also be used to probe the existence of dark matter \citep{Popolo2019}, investigate the potential of observable quantum gravity effects \citep{Yagi2016, Estes2017}, impose limits on the cosmological constant \citep{Iorio2018} and test for alternative theories of gravity \citep[e.g. scalar-tensor theories][]{Liu2014}. Evidently, there is therefore a huge scientific return from the detection and timing of a MSP-EMRB. \newline 

\noindent An alternative avenue for determining the system parameters of a massive BH is through gravitational radiation. The measurement of this radiation provides a complementary channel for precision tests of GR and astrophysics. Due to the time variation of the mass quadrupole moment of the gravitational field, a compact object in an elliptical orbit around a massive BH will emit a burst of radiation as it passes through periapsis \citep{Rubbo2006,Yunes2008,Berry2013}. This emission causes the orbit to slowly decay and circularize \citep{Peters1964}. Systems which continuously emit gravitational radiation and have an associated inspiral motion are know as Extreme Mass Ratio Inspirals \citep[EMRIs,][]{Babak2017,Berry2019}. These systems are a major class of target for the future observations by the next generation of space-based gravitational wave detectors such as the Laser Interferometer Space Antenna \citep[LISA,][]{Amaro2007}. EMRIs are particularly prized scientific targets for the same reasons as MSP-EMRBs; precision measurements in the strong-field regime, although in addition EMRIs inhabit a highly dynamical spacetime. \newline 

\noindent The typical orbital periods considered for MSP-EMRBs ($\sim$ 0.1 years) are not sufficiently short to continuously radiate gravitational radiation in the LISA frequency band ($\sim$ mHz). Moreover, the expected astrophysical rate of a `MSP-EMRI'  system is so low as to make a coincident, multimessenger, electromagnetic and gravitational detection unlikely \citep{Gair2017}. However, as a MSP in an EMRB passes through periapsis it will emit a burst of gravitational radiation.The coincident observation of the continuous electromagnetic MSP timing signal and the burst gravitational radiation from a MSP-EMRB offers a unique apparatus for multimessenger astronomy in strong-field environments. In addition, the continuous pulsar electromagnetic signal may aid in the detection of the accompanying gravitational radiation. Moreover, the burst gravitational radiation may in turn influence the received radio timing signal. Whilst burst waveforms are typically considered less informative than continuous EMRI waveforms, a multimessenger observation of a pulsar and GW burst may compound  the potential scientific return.  \newline 

\noindent In this work we explore the potential for the detection of burst gravitational radiation from typical MSP-EMRBs, of the sort typically considered  for precision tests of strong-field GR \citep[e.g.][]{Liu2012,Kimpson2019b}. We calculate the gravitational burst waveforms and signal-to-noise ratio (SNR) of a MSP-EMRB for generic orbits (i.e not restricted to e.g. equatorial plane, circular motion, spin-alignment). Firstly the waveforms are constructed via the semi-relativistic numerical kludge (NK) approach \citep{Babak2007,Berry2013MNRAS} where we  account for the relevant spin-spin, spin-orbit and spin-curvature couplings. With the waveforms determined, it is then possible to calculate the burst SNR from a MSP-EMRB. In addition, the GW emission might have an impact on the pulsar timing signal and so we explore the implications for the detection and modelling of MSP-EMRBs. \newline 

\noindent This paper is organized as follows. In Section \ref{sec:dynamics} we calculate the orbit of an extended spinning body around a spinning black hole, going beyond the point particle and geodesic approximation. In Section \ref{sec:waveforms} we then review the methods for mapping this orbital motion to the gravitational waveforms and determine the waveforms for typical MSP-EMRBs that are used as radio timing GR probes. In Section \ref{sec:SNR} we go on calculate the burst SNR of these systems for the most recent LISA configuration and noise model. In \ref{sec:PSRS} we briefly investigate the impact of a GW burst on the PSR timing signal. Discussion and concluding remarks are made in Section \ref{sec:disco}. \newline

\noindent We adopt the natural units, with $ c=G=\hbar = 1$, and a $(-,+,+,+)$ metric signature.  Unless otherwise stated, a c.g.s. Gaussian unit system is used in the expressions for the electromagnetic properties of matter. The gravitational radius of the black hole is $r_{\rm g} = M$ and the corresponding Schwarzschild radius is $r_{\rm s} = 2M$, where $M$ is the black-hole mass. A comma denotes a partial derivative (e.g.$\;\! x_{,r}$), and a semicolon denotes the covariant derivative (e.g.$\;\! x_{;r}$).

\section{Spin-Orbital Dynamics}
\label{sec:dynamics}
Test particles in GR follow geodesics, their motion determined by the background spacetime metric. However, clearly real astrophysical bodies such as PSRs are not test objects, but have a finite size and spin. Consequently, to properly describe the PSR dynamics one must consider higher order effects to account for the influence of the spin. In the extreme mass ratio, non-relativistic limit it is possible to describe the motion via two Hamiltonians \citep[e.g.][]{Iorio2012}. This describes the spin-spin and spin-orbit interaction and reproduces the effect of Lense-Thirring precession. However, additionally the spin of the pulsar will dynamically interact with the background spacetime curvature. In the presence of this spin-curvature coupling, the PSR motion is no longer a geodesic of the Kerr spacetime. Instead the PSR dynamical evolution can be described via the Mathisson-Papatrou-Dixon (MPD) formalism \citep{Mathisson1937,Papapetrou1951,Dixon1974}. \newline 

\noindent In general, the time evolution of the PSR is governed by,
\begin{eqnarray}
{T^{\mu \nu}}_{;\nu} = 0 \ ,
\end{eqnarray}
where $T^{\mu \nu}$ is the energy-momentum tensor. In the MPD formulation, a multipole expansion of the energy-momentum tensor is undertaken. This constructs the `gravitational skeleton'. In the extreme mass ratio limit, the pole and dipole terms are dominant and terms greater than the quadrupole can be neglected. The PSR dynamics are then entirely determined by the background gravitational field and the dynamical spin interaction with this field. The mass monopole is described by the 4-momentum $p^{\mu}$ whilst the spin dipole is $s^{\mu \nu}$. The corresponding equations of motion are \citep{Mathisson1937,Papapetrou1951,Dixon1974},
\begin{eqnarray}
\frac{D p^{\mu}}{d \tau} = - \frac{1}{2} {R^{\mu}}_{\nu \alpha \beta} u^{\nu} s^{\alpha \beta} \ ,
\label{Eq:mpd1}
\end{eqnarray}
\begin{eqnarray}
\frac{D s^{\mu \nu}}{d \tau} = p^{\mu} u^{\nu} -p^{\nu} u^{\mu} \ ,
\label{Eq:mpd2}
\end{eqnarray}
for proper time parameterization $\tau$ along the PSR worldline $z^{\alpha} (\tau)$, $D/d\tau $ denotes a covariant derivative whilst $u^{\nu}$ is the PSR 4-velocity and ${R^{\mu}}_{\nu \alpha \beta}$ the Riemann curvature tensor. In order to first construct the gravitational skeleton requires first specifying a reference worldline with which to define the expansion. Typically, such a choice would be the worldline described by the centroid (center of mass) of the body. However, in GR the centroid choice is observer-dependent. This uncertainty is evidenced the system of equations Eqs. \ref{Eq:mpd1} - \ref{Eq:mpd2}, which is not a determinate set since there exist more unknowns than equations. This is related to the uncertainty in choosing a reference world line for the multipole expansion. Explicitly choosing an observer with which with respect to which the centre of mass is defined renders the system of equations determinate. Such a choice is known as the spin supplementary condition (SSC). For this work we adopt the Tulczyjew-Dixon (TD) condition which specifies the centroid to be that measured in  zero 3-momentum frame:
\begin{eqnarray}
s^{\mu \nu} p_{\nu} = 0 \ ,
\label{Eq:ssc}
\end{eqnarray}
\citep{Tulczyjew1959,Dixon1964}. A key advantage of this SSC is that is specifies a unique worldline. In contrast, other choices of SSC are infinitely degenerate \citep[for discussion, including application to EMRIs, see][]{Costa2014, Babak2014}. With this choice of SSC both the mass of the MSP,
\begin{eqnarray}
m = \sqrt{- p^{\mu} p_{\mu}}
\end{eqnarray}
and the scalar contraction of the spin vector
\begin{eqnarray}
s = s^{\mu} s_{\mu}
\end{eqnarray}
are constants of the motion. In turn, the spin vector is a contraction of the spin tensor,
\begin{eqnarray}
s_{\mu} = -\frac{1}{2m} \epsilon_{\alpha \beta \mu \nu} p^{\nu} s^{\alpha \beta}
\end{eqnarray}

Since the black hole mass ($M$)  is much greater than pulsar mass ($m$), and the pulsar Moller radius $R_{\rm Moller}$ (the radius of the disk of all possible centroids) is much less than the radius of the pulsar, the pole-dipole terms are much stronger than the dipole-dipole terms. Therefore, to first order the 4-velocity and 4 momentum are parallel, i.e. $p^{\mu} \approx m u^{\mu}$. The equations of motion then become,
\begin{eqnarray}
\frac{D u^{\mu}}{d \tau} = - \frac{1}{2m} {R^{\mu}}_{\nu \alpha \beta} u^{\nu} s^{\alpha \beta} \ ,
\end{eqnarray}
\begin{eqnarray}
\frac{D s^{\mu \nu}}{d \tau} \approx 0 \ ,
\end{eqnarray} 
\citep{Chicone2005,Mashhoon2006}. 
The ordinary differential equations to then be integrated are \citep{Singh2005,Mashhoon2006}:
\begin{eqnarray}
\frac{dp^{\alpha}}{d\tau} = - {\Gamma^{\alpha}}_{\mu\nu} p^{\mu}u^{\nu} + \lambda \left( \frac{1}{2m} {R^{\alpha}}_{\beta \rho \sigma} \epsilon^{\rho \sigma}_{\quad \mu \nu} s^{\mu} p^{\nu} u^{\beta}\right) \ ,
\label{eq:MPD1}
\end{eqnarray}

\begin{eqnarray}
\frac{ds^{\alpha}}{d \tau} = - {\Gamma^{\alpha}}_{\mu \nu} s^{\mu}u^{\nu} + \lambda \left(\frac{1}{2m^3}R_{\gamma \beta \rho \sigma} \epsilon^{\rho \sigma}_{\quad \mu \nu} s^{\mu} p^{\nu} s^{\gamma} u^{\beta}\right)p^{\alpha} \ ,
\end{eqnarray}

\begin{eqnarray}
\frac{dx^{\alpha}}{d\tau} = -\frac{p^{\delta}u_{\delta}}{m^2} \left[ p^{\alpha} + \frac{1}{2} \frac{\lambda (s^{\alpha \beta} R_{\beta \gamma \mu \nu} p^{\gamma} s^{\mu \nu})}{m^2 + \lambda(R_{\mu \nu \rho \sigma} s^{\mu \nu} s^{\beta \sigma}/4)}\right] \ ,
\label{eq:MPD2}
\end{eqnarray}
where $s^{\mu}$ is the spin 4-vector and the dimensionless parameter $\lambda$ is used to label the terms which contribute to MPD spin-curvature coupling ($\lambda = 1$ includes spin-curvature coupling, for $\lambda = 0$ the coupling is omitted). In the $\lambda \rightarrow 0$ limit the conventional spin-spin and spin-orbit couplings are recovered. This set of ODEs are currently in a covariant form. Since we are concerned with the orbital dynamics around an astrophysical, spinning BH, we adopt the stationary and axisymmetric Kerr metric. In Boyer-Lindquist coordinates, the spacetime interval is then given by,
\begin{eqnarray}
{\rm d}s^2 = -\left(1 - \frac{2Mr}{\Sigma}\right) {\rm d}t^2 
- \frac{4aMr \sin^2 \theta}{\Sigma}\ {\rm d}t \;\! {\rm d}\phi 
+ \frac{\Sigma}{\Delta}{\rm d}r^2 + \Sigma\ {\rm d} \theta^2 \nonumber \\ 
\hspace*{1.2cm} + \frac{\sin^2 \theta}{\Sigma} \left[(r^2+a^2)^2 - \Delta a^2 \sin^2 \theta \right] {\rm d}\phi^2 \ , 
\label{eq:kerr_metric} 
\end{eqnarray}
where $\Sigma = r^2 + a^2 \cos^2 \theta$, $\Delta = r^2 - 2Mr +a^2$, $a$ is the black-hole spin parameter. With the metric specified, Eqs. \ref{eq:MPD1} - \ref{eq:MPD2} can be solved numerically via a 5th order Runge-Kutta-Fehlberg algorithm \citep{Press1977}. Going forward, we normalize the black hole mass $M=1$ and so the spacetime interval is then a lengthscale in terms of the gravitational radius $r_{\rm g} (=1)$.  We neglect the evolution of the orbital parameters due to the GW emission. The influence of GW burst emission on the orbital motion and a relativistic PSR timing model is discussed in Sec. \ref{sec:disco}.  
\subsection{Orbital specification}
\begin{figure}
	\includegraphics[width=\columnwidth]{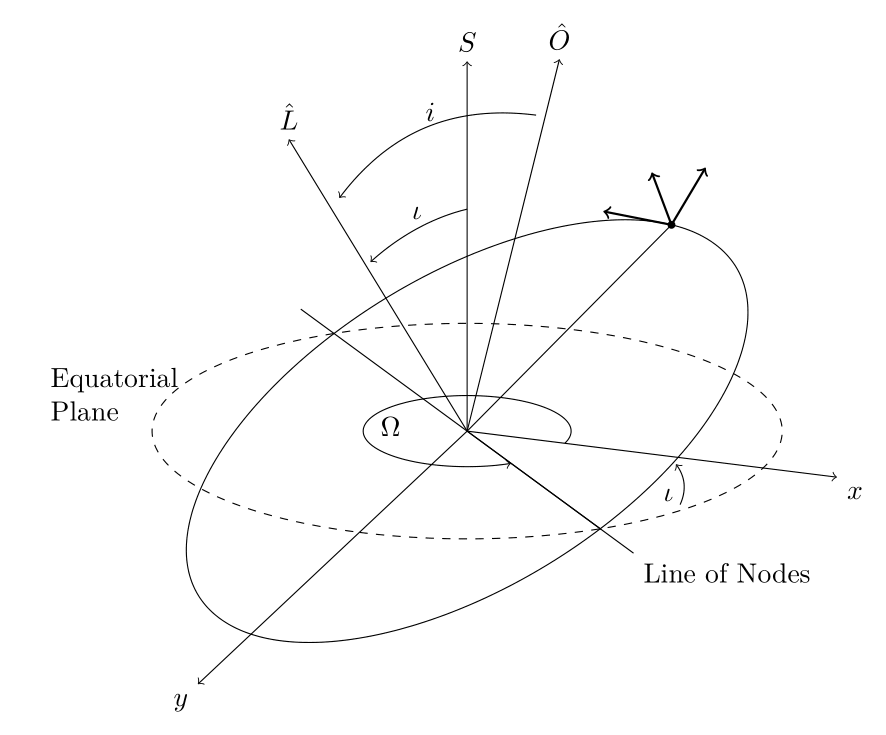}
	\caption{The coordinate system and orbital elements used to define the MSP motion. The orbital plane is inclined by an angle $\iota$ with respect to the BH equatorial plane - i.e. the $\theta = \pi/2$ plane about the BH spin axis. The inclination of the orbit with respect to the observer's line of sight is given by $i$ and the longitude of the ascending node as $\Omega$, where we have set the positive $x-$axis as the reference direction.}
	\label{fig:orbitalelements}
\end{figure}
The Kerr spacetime is stationary and axisymmetric and so is in possession of two Killing vectors,  $\xi^{t}, \xi^{\phi}$. These symmetries can then be related to conserved quantities. In particular the temporal Killing vector, $\xi^{t}$, is associated with the conservation of energy, $E$, whilst the azimuthal Killing vector is associate with the projection of the particle angular momentum along the black hole spin axis, $L_z$. In addition, the Kerr vacuum spacetime is well-known to harbor a rank-2 Killing tensor associated with a constant known as the Carter Constant, $Q$, \citep{Carter1968,DeFelice1999, Rosquist2009}. \newline 

\noindent For the purposes of astronomy, it is useful to map these constants to orbital parameters. In particular, the $E,L_z, Q$ can be mapped to the orbital elements $\mathcal{P},e, \iota$, the semi-latus rectum, eccentricity and inclination respectively. The semi-latus rectum and the eccentricity are effectively a reparameterization in turn of the apsidal approaches (periapsis distance $r_p$ and apoapsis $r_a$),
\begin{eqnarray}
r_p = \frac{\mathcal{P}}{1+e} \, \, \, ; \, \, \, r_a = \frac{\mathcal{P}}{1-e} \ ,
\end{eqnarray}
whilst the inclination angle $\iota$ is defined \citep{Glampedakis2002}
\begin{eqnarray}
\cos \iota = \frac{L_z}{\sqrt{Q+L_z}} \ .
\end{eqnarray}
We also define $i$ - as distinct from $\iota$- as the inclination of the orbit with respect to  the observer, i.e. 
\begin{eqnarray}
\cos i = \hat{L}_z \cdot \hat{O}
\end{eqnarray}
and $\hat{O}$ is the line of sight unit vector between the observer and the BH. The observer is at some distant radius  with polar and azimuthal coordinates $\Theta, \Phi$ respectively. The relevant geometry is described in Fig \ref{fig:orbitalelements}. The mapping between conserved quantities and orbital parameters is defined in \citep{Schmidt2002, Barausse2007} as 
\begin{eqnarray}
E=\sqrt{\frac{\kappa \rho+2 \epsilon \sigma-2 D \sqrt{\sigma\left(\sigma \epsilon^{2}+\rho \epsilon \kappa-\eta \kappa^{2}\right)}}{\rho^{2}+4 \eta \sigma}} \ ,
\end{eqnarray}
\begin{eqnarray}
L_{z}=-\frac{g_{1} E}{h_{1}}+\frac{D}{h_{1}} \sqrt{g_{1}^{2} E^{2}+\left(f_{1} E^{2}-d_{1}\right) h_{1}} \ ,
\end{eqnarray}
\begin{eqnarray}
Q=z_{-}\left[a^{2}\left(1-E^{2}\right)+\frac{L_{z}^{2}}{1-z_{-}}\right] \ ,
\end{eqnarray}
where
\begin{eqnarray}
z_{-} = \sin^2 \iota
\end{eqnarray}
and $D = \pm 1$ denotes prograde and retrograde orbits.  In turn, the functions are defined as,
\begin{eqnarray}
f(r) \equiv r^{4}+a^{2}\left[r(r+2)+z_{-} \Delta\right] ' ,
\end{eqnarray}
\begin{eqnarray}
g(r) \equiv 2 a r \ ,
\end{eqnarray}
\begin{eqnarray}
h(r) \equiv r(r-2)+\frac{\Delta z_{-}}{1-z_{-}} \ ,
\end{eqnarray}
\begin{eqnarray}
d(r) \equiv \Delta \left(r^{2}+a^{2} z_{-}\right) \ ,
\end{eqnarray}
and for the eccentric orbits considered in this work,
\begin{eqnarray}
(f_1, g_1, h_1, d_1) = f(r_p), g(r_p), h(r_p), d(r_p) \ , \\
(f_2, g_2, h_2, d_2) = f(r_a), g(r_a), h(r_a), d(r_a) \  ,
\end{eqnarray}
and
\begin{eqnarray}
\kappa & \equiv d_{1} h_{2}-d_{2} h_{1} \ , \\
\varepsilon & \equiv d_{1} g_{2}-d_{2} g_{1} \ ,\\
\rho & \equiv f_{1} h_{2}-f_{2} h_{1}  \ ,\\ 
\eta & \equiv f_{1} g_{2}-f_{2} g_{1} \ , \\
\sigma & \equiv g_{1} h_{2}-g_{2} h_{1} \ .
\end{eqnarray}
Naturally this approach of mapping conserved quantities to Keplerian orbital elements is only an approximation and does not account for variations in $L$ or $e$ over the orbit due to spin-curvature coupling \citep[e.g. ][]{Singh2014}. However for our purposes it will prove sufficient and a useful approximation to describe the sorts of orbits that we want to model. 
\section{Constructing the Waveforms}
\label{sec:waveforms}
The generation of sufficiently accurate waveforms from compact objects around a massive BH companion is currently a major enterprise in order to realise the scientific potential of LISA-EMRIs \citep[e.g.][]{Chua2015,Meent2017,Pound2017,Barack2019}. Via perturbation theory in the extreme mass ratio limit, waveforms accurate to first-order can be calculated accounting for the self-force or back-reaction effects of the GW radiation on the orbit. In order to accurately track the orbit over the large number of cycles that are expected to be observable with ERMIs, calculations accurate to second order are required and work in this area is ongoing \citep{Pound2017PhRv,Moxon2018}. Given both the theoretical complexity and computational cost of calculating consistent waveforms in this way, alternative fast yet accurate models have been developed by a number of authors \citep[e.g.][]{Barack2004,Babak2007, Chua2017}. For our purposes, we adopt the Numerical Kludge (NK) approach of \citet{Babak2007}. The GW `recipe' within the NK framework has two primary ingredients. First the orbital trajectory of the object is specified. Typically the motion is described as that of a test body following a geodesic on a Kerr background spacetime. However, as discussed this neglects the extended nature of real astrophysical pulsars and the associated spin couplings. For this work we go beyond the point-particle geodesic approximation and consider generic orbits of extended objects around a Kerr BH, specifying the orbital trajectory via the MPD equations outlined in the preceding section which properly account for the dynamical spin effects. Once the orbital motion has been specified, the Boyer-Lindquist coordinates of the background curved spacetime are mapped to flat-space spherical polar coordinates. The waveform can then be constructed from the well-known expressions for gravitational waves from flat-space trajectories. \newline 

\noindent Naturally the NK approach is not self-consistent; the gravitational radiation is generated assuming a flat background spacetime whilst the orbital motion instead assumes a curved Kerr geometry (with associated relativistic spin couplings). However it does exhibit a remarkable agreement with the more computationally expensive perturbative methods. Specifically, the overlap between the NK and the more intensive, accurate waveforms is $ > 95 \%$ across most of the parameter space. \citet{Babak2007} offer the rule of thumb that NK waveforms are appropriate as long as the periapsis distance is greater than $\sim 5 r_g$. Since we are concerned with pulsars on typical orbits that will be used for radio timing tests of GR, we will exclusively deal with orbits $r_p > 5 r_g$. However, it is worth noting that this rule of thumb was considered for orbits which continually emit in the LISA frequency range and so need to be tracked over a large number of cycles. For single bursting passages it seems reasonable that the NK approach could be pushed below this limit.The numerical kludge approach is also  applied to eccentric orbits in \citet{Berry2013MNRAS}. 

\subsection{Time domain waveforms}
Within the NK framework the usual`plus', $h_+$, and `cross', $h_{\times}$, polarisations of the gravitational wave in the  time-domain are given by,
\begin{eqnarray}
h_+ = h^{\Theta \Theta} - h^{\Phi \Phi} \ ,
\end{eqnarray}
\begin{eqnarray}
h_{\times} = 2 h^{\Theta \Phi} \ ,
\end{eqnarray}
where $h^{\Theta \Theta}, h^{\Phi \Phi}, h^{\Theta \Phi}$ are functions of the 
polar and azimuthal coordinates of the observer ($\Theta, \Phi$ respectively) and the GW perturbation ($h_{\mu \nu}$) on the background Minkowski spacetime. The trace-reversed metric perturbation is given to octupole order as\citep{Bekenstein1973,Press1977}
\begin{eqnarray}
\bar{h}^{jk} = \frac{2}{r} \left[ \ddot{I}^{jk} - 2 n_i \ddot{S}^{ijk} + n_i \dddot{M}^{ijk}\right] \ ,
\label{hEQ}
\end{eqnarray}
where $n_i$ is the radial unit vector pointing to the observer and $I^{jk}$, $S^{ijk}$, $M^{ijk}$ are the mass quadrupole, current quadrupole and mass octupole respectively. In order to calculate the second and third derivatives of the multipole moments necessary for calculating Eq. \ref{hEQ} we use a numerical finite difference scheme \citep{Fornberg}. \newline 

\noindent Example burst time-domain waveforms from a pulsar going through periapsis at the Galactic centre are presented in Figure \ref{fig:example orbits}. Variations in the orbital parameters naturally influence the resultant waveform; e.g. more eccentric orbits have a more localised burst with greater amplitudes whilst for more circular orbits the waveform is more extended. This will have implications for the SNR and detectability of GW bursts from MSP-EMRBs. The target pulsars have typical orbital periods of $\sim 0.1$ year and so over a year of observation there will be multiple GW bursts as the PSR repeatedly passes through periapsis, as presented in Fig. \ref{fig:com}. As the observer angle changes, so too does the wave amplitude, particularly for the $h_{\times}$ mode.

\begin{figure}
	\includegraphics[width=\linewidth]{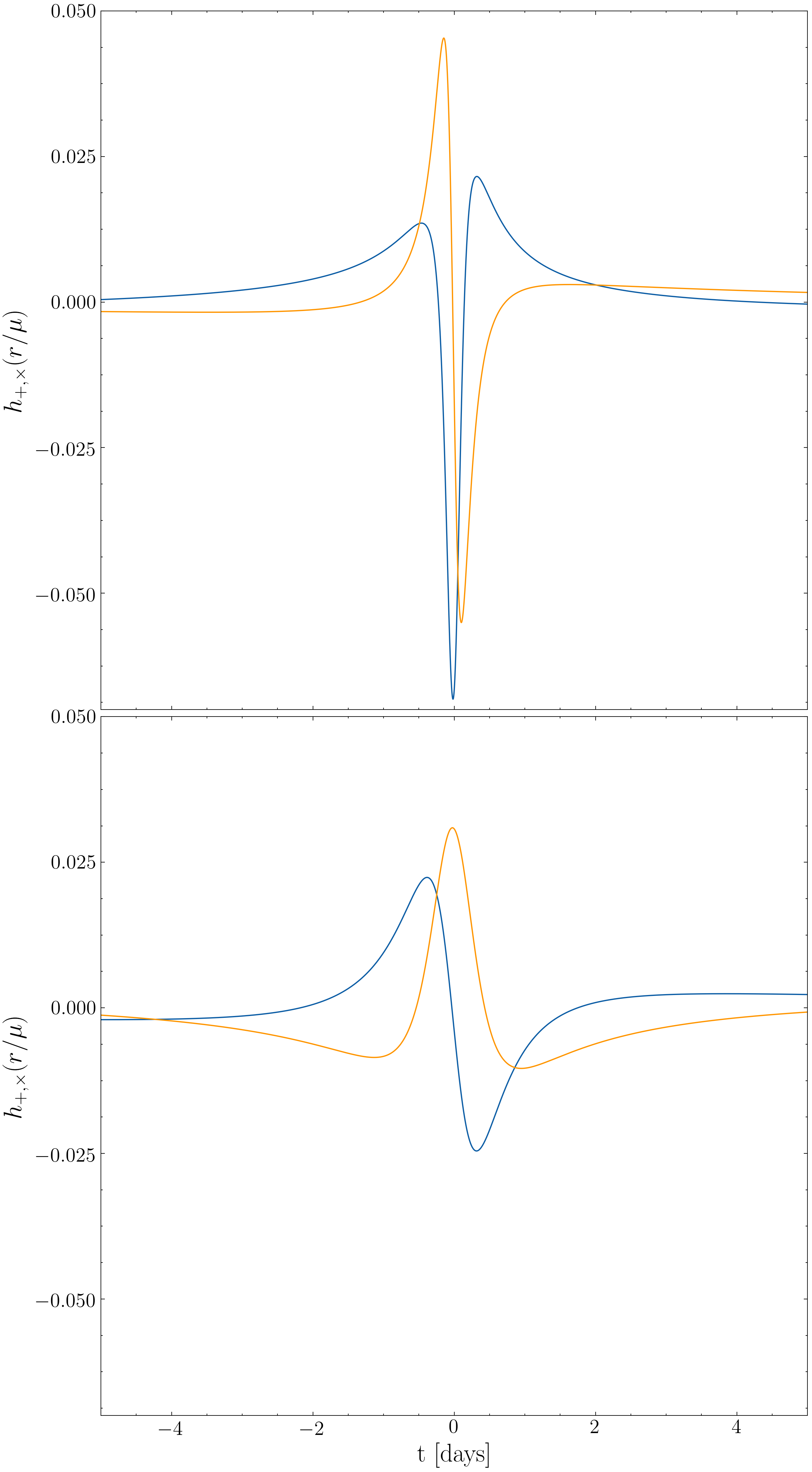}
	\caption{\textit{Top panel:} An example gravitational burst waveform from a $1.4 M_{\odot}$ mass object with orbital parameters $a=0.6$, $e=0.9$, $P = 0.1$ years, $\iota = 15$ degrees, $\Omega = 0$ radians and the observer is situated at $\Theta = \Phi= 0$. The BH has mass $4.3 \times 10^6 M_{\odot}$ and spin parameter $a=0.6$. Both the $+$ (blue) and $\times$ (orange) polarisations are presented. \textit{Bottom panel:} As top panel but for $a=0.85$, $e=0.8$, $\iota = 25$, $\Omega = 3\pi/4$. The observer angles and orbital period are unchanged from the first case. The waveform is more extended compared to the more eccentric case, and the amplitude is reduced.} \label{fig:exampleorbitsa}
\end{figure}

\begin{figure*}
	\subfloat[\label{fig:example_trajecotry}]{\includegraphics[width=0.48\textwidth]{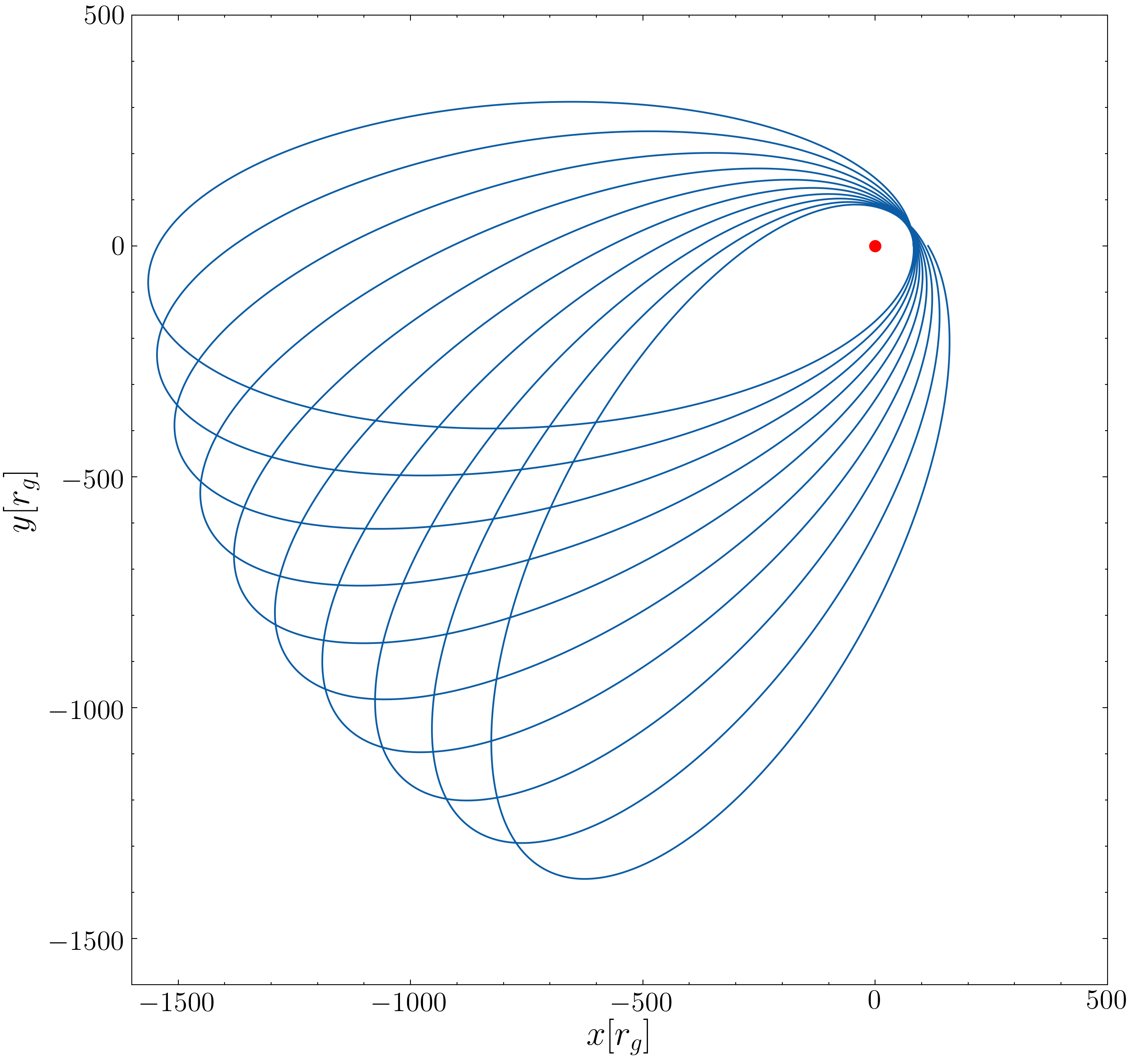}}
	\subfloat[\label{fig:example_waveform}]{\includegraphics[width=0.48\textwidth]{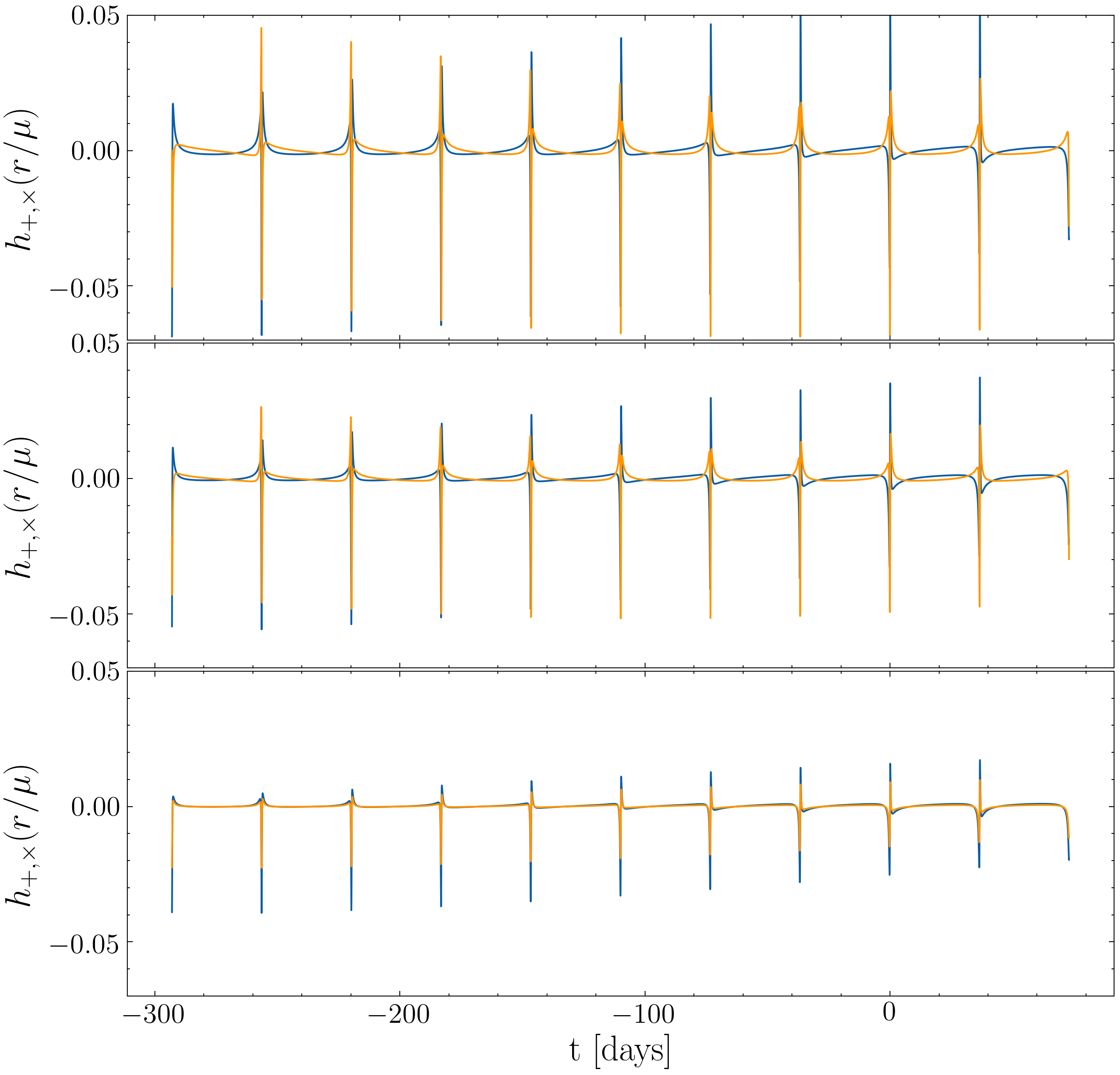}}
	\caption{Orbital system of a 1.4 $M_{\odot}$ mass MSP with orbital parameters $a=0.6$, $e=0.9$, $P = 0.1$ years, $\iota = 15$ degrees, $\Omega = 0$ radians (i.e. as in Fig. \ref{fig:example orbitsa}), observed over over 10 orbits (i.e. 1 year observation) \textit{(a) }Orbital trajectory in the $x-y$ plane a around a spinning black hole. Precession of apsis is clearly visible. \textit{(b)} Waveforms for $\Theta = 0,\pi/4, \pi/2$ (top,middle,bottom panels respectively). The waveforms amplitude decreases with increasing $\Theta$.} \label{fig:com}
\end{figure*}
\begin{figure}
	\subfloat[\label{fig:example orbitsa}]{\includegraphics[clip,width=\columnwidth]{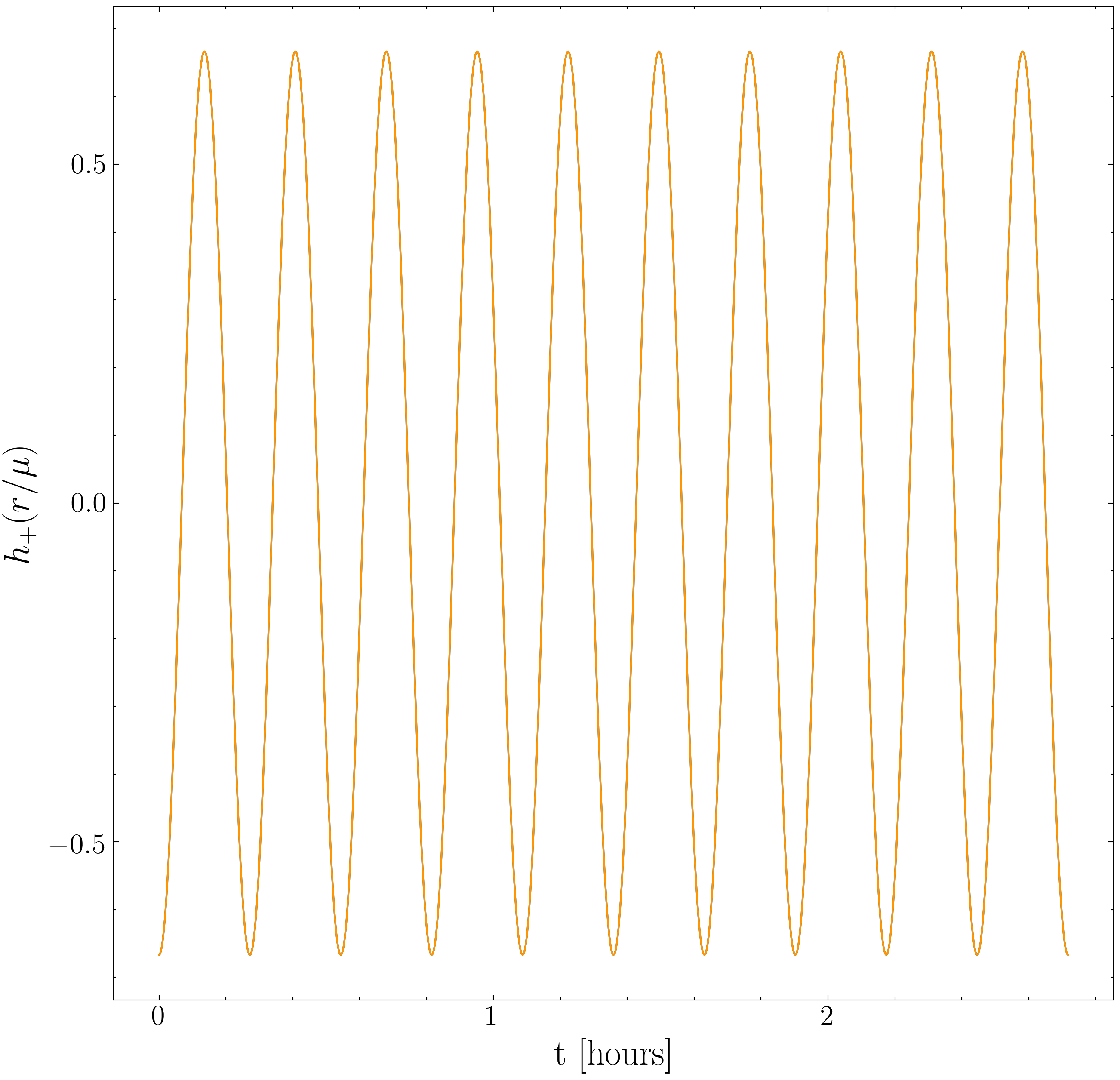}}	
	\caption{The gravitational waveforms in the time domain from equatorial circular orbital motion  ($r = 6 r_g$ and $\Theta = 0$) around a non-spinning ($a=0$, Schwarzchild) BH. Both the NK and analytical solutions are presented, but cannot be resolved due to the high degree of overlap $\mathcal{O} = 0.9999999999999997 $.} \label{fig:overlap}
\end{figure}

Before proceeding further, as a sanity check it is desirable to compare our numerical kludge waveforms with some accurate, analytical template. For the specific case of circular and equatorial orbits, in the distant limit, it is possible to derive exact analytical expressions for the $h_{+, \times}$ waveforms \citep{Gourgoulhon2019}. Turning `off' the relativistic spin couplings of the MPD formalism, we can then also describe a circular, equatorial orbital geodesic numerically and generated the associated waveforms. Following \citet{Babak2007,Chua2017},we define the overlap between two waveforms as,
\begin{eqnarray}
\mathcal{O} = (\hat{a}|\hat{b})
\end{eqnarray}
where $\hat{a}$ denotes a normalized unit vector such that and $(\hat{a}|\hat{b})$ denotes an inner product. The inner product is defined with respect to the noise power spectral density $P_n(f)$, which are both described in the next section. Identical waveforms have $\mathcal{O} = 1$ whilst for completely anti-correlated signals $\mathcal{O} = -1 $ and $\mathcal{O} = 0$ for orthogonal signals. To check the accuracy of our MPD + NK approach, we take the circular, equatorial orbit with $r = 6 r_g$ and $\Theta = 0$ \citep[c.f. Fig 3 of][]{Gourgoulhon2019} around a non-spinning ($a=0$) BH and generate the gravitational waveform numerically. The trajectory and waveform are presented in Figure \ref{fig:overlap}. To the eye the numerical and analytical waveforms are completely overlaid and cannot be resolved. More quantitatively, in double precision $\mathcal{O} = 1-3 \times 10^{-16}$. i.e. the waveforms are identical to machine precision. This is especially encouraging since the $r = 6 \, r_g$ regime around the BH is explicitly the strong field regime. 

\section{Signal to Noise ratio}
\label{sec:SNR}
In order to establish whether the burst gravitational radiation from MSP-EMRB system is detectable, it is necessary to calculate the burst signal to noise ratio (SNR), $\rho$. The general inner product is defined,
\begin{eqnarray}
(a|b) = 2 \int^{\infty}_0 \frac{\tilde{a}^*(f)\tilde{b}(f)+\tilde{b}^*(f)\tilde{a}(f)}{P_n(f)} df \ ,
\end{eqnarray}
$f$ is the frequency and $P_n(f)$ is the noise power spectral density \citep[PSD, ][]{Cutler1994}. 
The gravitational wave signal recorded by the detector is a linear combination of the two polarisation modes, corrected for the the response functions $F_{+, \times}$ of the LISA instrument,
\begin{eqnarray}
\tilde{h}(f) = F_+(\Theta, \Phi, \Psi, f) \tilde{h}_+(f) + F_{\times}(\Theta, \Phi, \Psi, f) \tilde{h}_{\times}(f)
\end{eqnarray}
where $\Psi$ is the polarisation angle. 
Denoting the sky and polarisation average as $\langle \rangle$, the averaged GW signal is given by
\begin{eqnarray}
\langle \tilde{h}(f) \tilde{h}^*(f) \rangle = \mathcal{R}(f) \left( |\tilde{h}_+(f)|^2 + \tilde{h}_{\times}(f)|^2  \right)
\end{eqnarray}
where $\mathcal{R}$ is the instrument response function averaged over the sky ($\Theta, \Phi$) and polarization angle ($\Psi$): 
\begin{eqnarray}
\mathcal{R} (f) = \langle F_{+} (f) F_{+}^*(f) \rangle =\langle F_{\times} (f) F_{\times}^*(f) \rangle
\end{eqnarray}
see \citet{Robson2019} for details. Consequently, the effective SNR used in this work is given by,
\begin{eqnarray}
\rho^2 = 4 \int^{\infty}_0 \frac{|\tilde{h}_+(f)|^2 + |\tilde{h}_{\times}(f)|^2}{S_n(f)} df
\label{eq:snr}
\end{eqnarray}
where 
\begin{eqnarray}
S_n(f) = \frac{P_n(f)}{ \mathcal{R}(f)}
\label{eq:noise}
\end{eqnarray} 
This definition is also used for the study of GW from Galactic Centre objects in \cite{Gourgoulhon2019}.

\subsection{LISA Noise Model}
\label{sec:LISAnoise}
The instrument response function does not have a closed form expression, but can be well fit as \citep{Robson2019},
\begin{eqnarray}
\mathcal{R}(f) = \frac{3}{10} \frac{1}{1+0.6(f/f_*)^2} \ ,
\end{eqnarray}
and $f_*$ is the LISA transfer frequency. However, instead of this form we use the exact response function as given by \cite{RobsonGIT2018}. The LISA noise PSD is given by
\begin{eqnarray}
P_n(f) = \frac{P_{\rm OMS}}{L^2} + 2(1+\cos^2(f/f_*)) \frac{P_{\rm acc}}{(2\pi f)^4 L^2} \ ,
\end{eqnarray}
for LISA arm length $L $. The  optical metrology noise,
\begin{eqnarray}
P_{\rm OMS} = (1.5 \times 10^{11} \text{ m})^2 \left(1 + \left(\frac{2 \text{ mHz}}{f}\right)^4\right) \text{ Hz}^{-1} \ ,
\end{eqnarray}
and the acceleration noise is,
\begin{align}
P_{\rm acc} = (3 \times 10^{-15} \text{ m s}^{-2})^2 \left(1 + \left(\frac{0.4 \text{ mHz}}{f}\right)^2\right) \nonumber \\ 
\left(1 + \left(\frac{f}{0.4 \text{ mHz}}\right)^4\right) \text{ Hz}^{-1} \ .
\end{align}
In addition to the instrumental noise, there is also an additional non-stationary noise contribution from the population of compact galactic binaries. This noise can be well described by the parametric function \citep{Cornish2017}
\begin{align}
S_c(f) = A f^{-7/3} e^{-f^{\alpha} + \beta f \sinh(\kappa f)} \left[ 1 + \tanh (\gamma (f_k - f))\right] \text{Hz} ^{-1}
\end{align}
For our bursting sources we use with fit parameters relevant for observation times less than 6 months, given in \citet{Robson2019} The characteristic strain is defined,
\begin{eqnarray}
h_c^2 = f(S_n(f) + S_c(f)) 
\label{eq:strain}
\end{eqnarray}

\subsection{Windowing}
In transforming the GW signal from the time domain to the frequency domain we are necessarily performing a Fourier transform on a finite signal. As a consequence the signal in the frequency space exhibits spectral leakage; extra components in the frequency regime, due to the fact that the time series is not exactly zero valued at the edges of the time interval $T$ over which the Fourier transform takes place. \newline

\begin{figure*}
	\includegraphics[width=\textwidth]{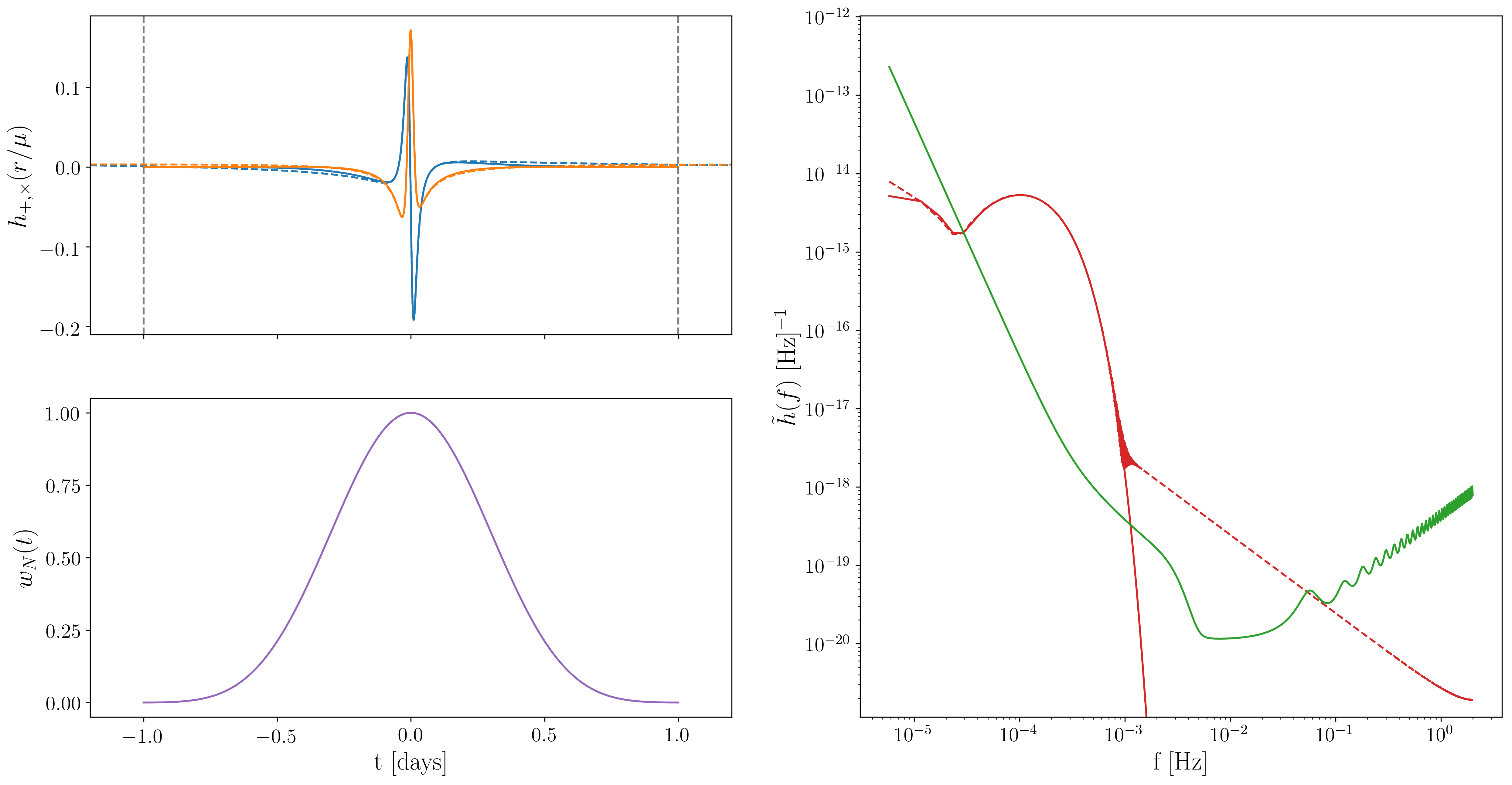}
	\caption{\textit{Left panel, top:} The time domain gravitational waveform from a MSP orbiting the Galactic Centre BH with orbital period $P = 0.01$ years, $e=0.9$, $\iota = 30 \deg$, $\Omega = \pi/2$, $a=0.6$, $\Theta = \Phi = 0$. The `$+$' (blue) and `$\times$' (orange) polarisations are shown, with no window (coloured, dashed) and after being windowed by the Nuttal function (coloured, solid). The vertical grey dashed lines indicate the width of the Nuttal window. \textit{Left panel, bottom:} The Nuttal window in the time domain. \textit{Right panel:}  The LISA noise curve (green) $\sqrt{S_n(f)}$ (not the characteristic strain $h_c$, Eq. \ref{eq:strain}) as described by Eq. \ref{eq:noise}. The red lines are the waveform in the frequency domain $\sqrt{\tilde{h}_+(f)^2 + \tilde{h}_{\times}(f)^2}$ both with windowing (solid) and without (dashed).} \label{fig:windowing}
\end{figure*}

\noindent To counter the effects of spectral leakage we first multiply our GW time series with a window function which tapers the signal to zero outside of the interval $T$.  For this work we adopt the Nuttal window with continuous first derivative \citep{Nuttal}:
\begin{eqnarray}
w[n] = a_0 -a_1 \cos \left( \frac{2 \pi n}{N}\right) + a_2 \cos \left( \frac{4 \pi n}{N}\right) - a_3 \cos \left( \frac{6 \pi n}{N}\right)
\end{eqnarray} 
where $N$ is the window length, $0 \leq n \leq N$ and $(a_0,a_1,a_2,a_3) = (0.355768,0.487396,0.144232,0.012604)$. Whilst other choices of window function are available, the Nuttal window is both computationally inexpensive to evaluate and exhibits good performance for the parameter space explored in this work \citep[see e.g. Appendix A of ][]{Berry2013MNRAS}. The effects of the Nuttal window on the frequency spectra of a particular MSP-BH system are presented in Fig. \ref{fig:windowing}

\subsection{SNR of a Astrophysical MSP-EMRB}
We are now in a position to calculate the GW SNR from a typical radio MSP-EMRB. We are concerned here with MSP-EMRB systems which have orbital parameters which render them useful from the perspective of testing strong-field GR via radio PSR timing \citep[e.g.][]{Liu2012, Psaltis2016, Kimpson2019b}. Typically the principal search area discussed for these systems is the Galactic centre. Whilst this region is an important target, globular clusters and dwarf spheroidal galaxies provide alternative, cleaner search grounds, without the typical problems of the Galactic centre c.f. scattering and dispersion \citep[e.g.][]{Rajwade2017,Kimpson2019a}. With the exception of the Small and Large Magellanic clouds, no extragalactic pulsars have currently been detected \citep{Noori2017}, nor have we detected MSPs sufficiently close to Sgr A* for appropriate tests of GR. However, with the advanced sensitivity of the next generation of radio telescopes (e.g. FAST/SKA), and the development of more sophisticated search algorithms - appropriate for the strong-field, relativistic environments that these systems inhabit - there is a real possibility to detect and time MSPs both in the Galactic centre and in external regions (e.g. globular clusters). Indeed, it is expected that SKA will be able to detect PSRs of the Local Group \citep{Keane2015}. Moreover, long integrations of a specific target - in contrast to wide field search sky imaging -  may also allow these systems to be detected. It is therefore prudent to consider not just the GC, but also nearby globular clusters and dwarf spheroidal galaxies. We now explore the GW SNR from MSP-EMRBs in each of these environments in turn.
\subsubsection{Galactic Centre}
Typically studies looking at using MSP-EMRBs to test GR have focused on systems at the Galactic centre. Due to the proximity of the central massive BH this target is highly appealing, but for strong-field GR tests there are some requirements on the sort of orbit. To avoid external Newtonian perturbations \citep{Merritt2011} contaminating a `clean' GR test, it is necessary for consistent timing campaigns to observe orbits with $P < 0.1$ years \citep{Liu2012}. Alternatively, dense timing campaigns close to periapsis \citep{Psaltis2016} of some longer, eccentric orbit may also mitigate these external perturbations. MSP-EMRBs are expected to retain significant eccentricities since during their formation they are scattered by two body interactions \citep[see e.g. ][for a comprehensive review of the formation mechanisms of extreme mass ratio binaries]{Amaro2011,Amaro2018} or else supernova kicks \citep[e.g.][]{Bortolas2019} on to eccentric orbits, and have not had sufficient time to circularize due to GW emission. Eccentric orbits are highly appealing since the periapsis approach scales with the eccentricity as 
\begin{eqnarray}
r_p \propto \frac{1}{1+e} \ ,
\end{eqnarray}
and so more eccentric orbits will probe stronger gravitational fields, as well as being less susceptible to external Newtonian perturbations. Eccentric orbits are similarly useful for GW bursts, since the GW burst strength scales with the periapsis approach \citep{Berry2013}. \newline

\begin{figure}
	\includegraphics[width=\linewidth]{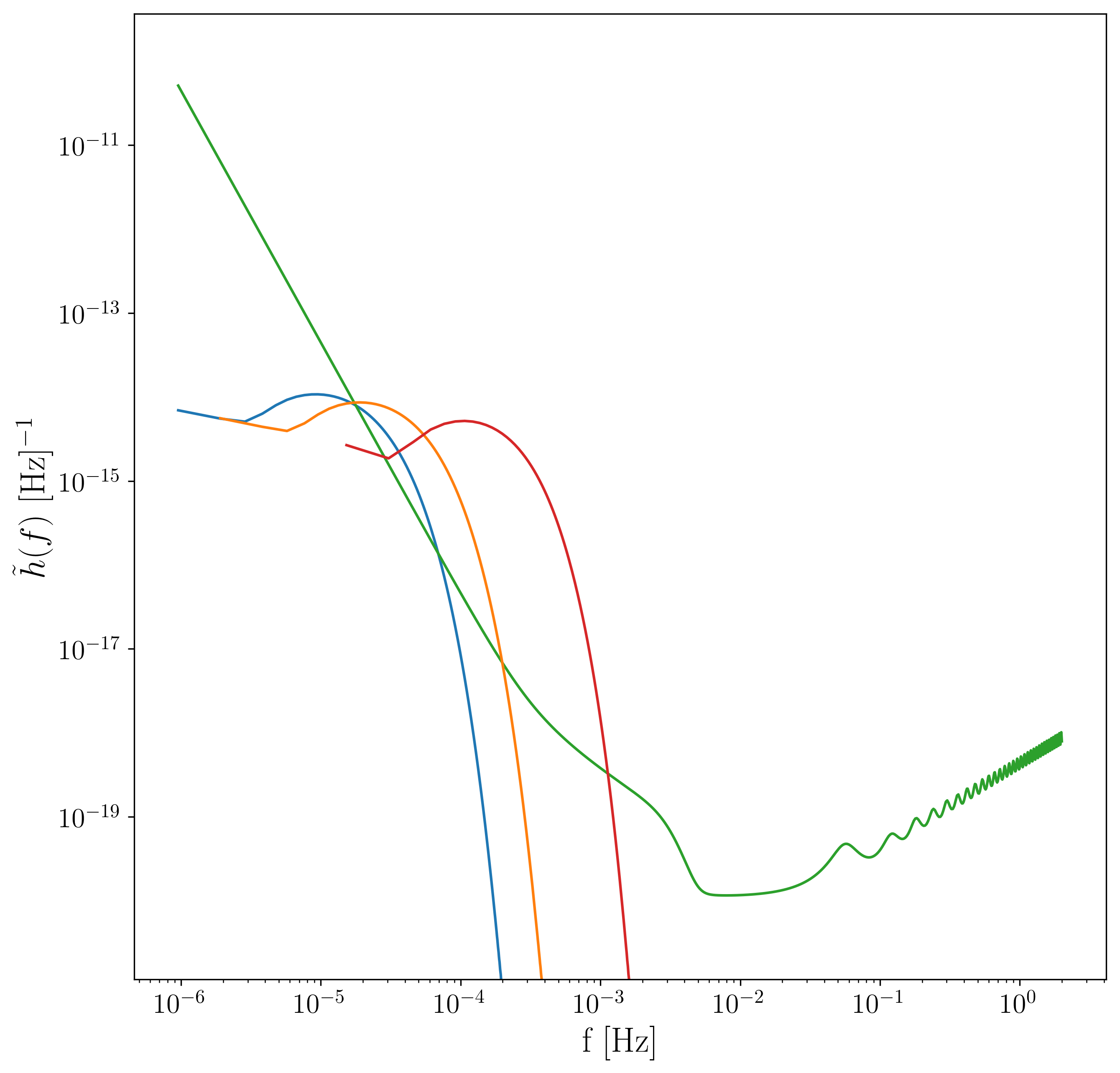}
	\caption{The frequency spectra of 3 MSP-EMRB systems at the Galactic centre with eccentricity $e=0.9$ and orbital periods $P = 0.01, 0.05,0.1$ (blue, orange,red respectively). The inclination angle is $\iota = 30 \deg$ and we have set the MSP mass to be $1.4 M_\odot$. The SNR of each of these systems is $\rho = 22, 0.25, 0.03$ respectively.} \label{fig:multispectra}
\end{figure}

\begin{figure}
	\includegraphics[width=\linewidth]{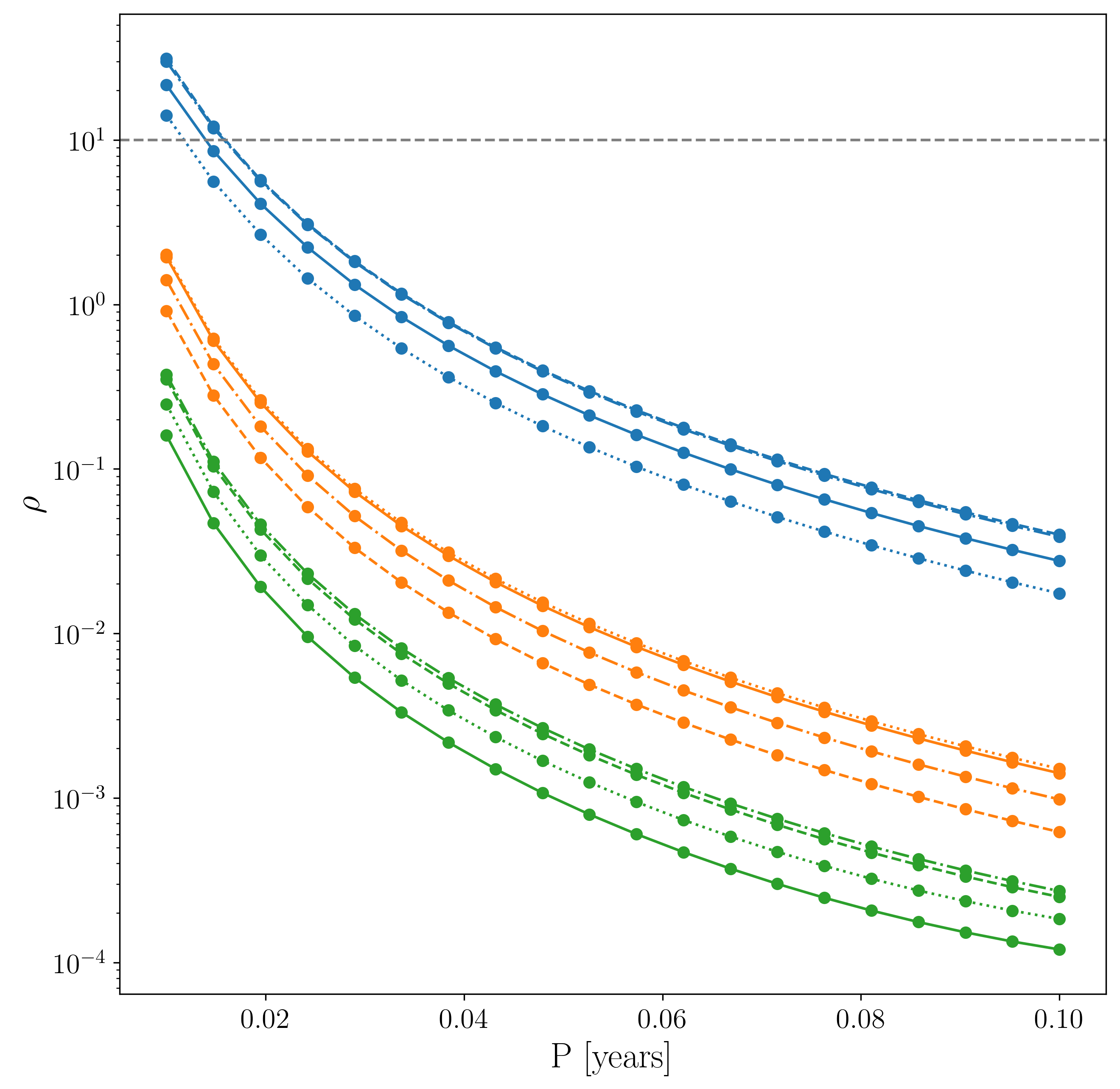}
	\caption{The signal to noise ratio of Galactic centre MSP-BH systems at $e=0.9,0.8,0.7$ (blue, orange,green respectively) at inclinations $\iota = 0,30,60,80 \deg$. More eccentric orbits with shorter orbital periods have greater values of $\rho$, whilst the SNR is also weakly dependent on how face on the orbit is to the observer. The grey horizontal line denotes $\rho = 10$, a typical cutoff for the minimal detection SNR in GW astronomy. However, given the presence of an accompanying EM beacon, the detection could be pushed below this threshold.} \label{fig:SNRparam}
\end{figure}

\noindent For MSP-BH systems at the Galactic centre we set the Galactic centre BH mass to be $4.31 \times 10^6 M_{\odot}$ and at a distance $8.33$ kpc \citep{Gillessen2009}, with spin parameter $a=0.6$. The spectra of 3 Galactic centre MSP-EMRB systems with orbital periods $P = 0.01, 0.05, 0.1$ years and $e=0.9$ are presented in Fig. \ref{fig:multispectra}. The MSP mass is set to be $1.4 M_{\odot}$, whilst we have set $\iota = 30 \deg$, and the longitude of the ascending node to be $\Omega = \pi/2$, with the observer located in the Galactic plane at $\Theta = \pi/2, \Phi  = 0$, assuming the spin axis of the central black hole is perpendicular to the Galactic plane (see Fig. \ref{fig:orbitalelements}).
It is immediately evident, as expected, that those systems with shorter orbital periods and hence closer periapsis passages will exhibit stronger signals. Decreasing the orbital period from $P = 0.1$ years to $P=0.01$ causes a corresponding shift in the frequency spectra to the region where LISA has greatest sensitivity (towards $\sim$ mHz). The SNR from each of these systems is $\rho = 22, 0.25, 0.03$ for $P= 0.01, 0.05,0.1$ years respectively. The SNR will be influenced not only by the orbital period, but also secondary factors such as the eccentricity, system orientation with respect to the BH and the observer ($\iota, \Theta$) and the BH spin. The exploration of this parameter space is shown in Fig. \ref{fig:SNRparam}. It can be seen that the major factors which influence the SNR are the orbital period and the eccentricity, with the influence of secondary factors such as system orientation or BH spin causing smaller deviations. As expected, short period, highly eccentric systems corresponds to the greatest SNR values. Typically in GW data analysis, for a system to be detectable, it requires that the SNR is greater than $\sim 10$. For this limit, only those systems at the Galactic centre with orbital periods less than $\sim 0.02$ and eccentricities $e \sim 0.9$ years might emit gravitational burst radiation which is detectable.
\noindent The exact physical interpretation of SNR here warrants further exploration. The output of a GW detector like LIGO or LISA consists of a superposition of the detector noise $n(t)$ and the gravitational wave signal $h(t)$,
\begin{eqnarray}
s(t) = h(t) + n(t)
\end{eqnarray}
The aim of GW data analysis is then to try and extract $h(t)$. In standard, blind GW astronomy (i.e. no continuous EM counterpart) for continuously radiative sources (e.g. LIGO/Virgo BH-BH mergers, LISA EMRIs) high values of $\rho$ are desirable since this governs both the probability of detection and the precision with which the system parameters can ultimately be determined. In order to maximise the SNR and to combat the weak instantaneous GW signal, typical GW sources are observed for a large number of cycles to allow the signal to build up above the noise. LISA EMRIs in particular are expected to spend a large number of cycles radiating in the LISA frequency band; 1 year of observation would be required to detect a $1 M_{\odot}$ object in a circular orbit at $r=50$ around a Schwarzschild BH at the Galactic centre \citep[Fig. 7,][]{Gourgoulhon2019}. Typical bursting sources with large amplitudes such as cosmic strings \citep[e.g.][]{Aasi2014} are searched for by simply looking for an excess of power, which shows above the instrument noise. Eccentric compact objects orbiting the Galactic centre typically do not have a continuous inspiral emission, but are instead bursting sources. Consequently, these systems are only detectable if the bursting GW amplitude is sufficiently large, requiring a large mass compact object with a close periapsis passage. As a result, BHs with masses of 10s of $M_\odot$ at 10s of $r_g$ are typically considered as the primary sources, whilst the detection of signals from lower mass ($\sim 1 M_\odot$) pulsars or white dwarfs with longer orbital periods is less likely. \newline

\noindent However, MSPs acting as a BH companion and a GW source provide a unique advantage over other types compact objects (white dwarfs, BHs) in that they have a continuous electromagnetic beacon. As a consequence, if a pulsar is observed via radio timing around the Galactic centre, it would be possible to derive a precision prediction of the expected gravitational waveform and the expected time at which this waveform would be received by an observer. Electromagnetically dark bursting waveforms are typically poorly modeled and not amenable to matched filtering methods, whilst GW burst from MSP sources would have exceptionally well modeled waveforms. With this information known apriori it would be possible to detect GW signals via matched filtering methods which is usually not possible for dark bursting sources. As a consequence, whilst the SNR  as given by Eq. \ref{eq:snr} are a useful standard metric for quantifying the strength of the GW signal, since the expected signal is well-known apriori it may prove possible to detect bursting signals at SNRs that are typically though of as being undetectable. Once the waveform has been detected using a EM-informed template, one could then further refine the waveform model to match the observations for parameter estimation. Clearly there will be some lower limit at which point the noise is completely dominant over the signal and no GW can be detected, even when the expected signal is known. However, for the Galactic centre pulsars that are typically considered for radio timing tests of GR these signals should be detectable via matched filtering, even for longer orbits. \newline

\subsubsection{Stellar Clusters}
In addition to the Galactic centre, there exist other potential hunting grounds for MSP-EMRBs which can be used via radio timing for strong-field tests of GR. As discussed, consilient strands of evidence suggest that IMBH could reside in globular clusters \citep{Mezcua2017}, although some of this evidence is disputed and definitive  `smoking-gun' evidence for IMBH is still lacking. Nevertheless, if the centre of globular clusters do host IMBH, then these regions would be ideal places to search for MSP-EMRBs without the additional complications raised due to scattering of the MSP radio pulse, as is expected for observations of the Galactic centre. The centre of globular clusters have remarkably high stellar densities \citep[$\sim 10^6$ stars per cubic parsec, ][]{Freire2013} and indeed the number of pulsars per unit mass is a factor of $10^3$ greater than in the Galactic disk on account of these high stellar densities allowing dynamical effects like mass segregation to be more efficient. Since MSPs are thought to evolve from low mass X-ray binaries (LMXBs) and globular clusters are known hosts of abundant LMXB populations, MSPs make up a significant fraction of the globular cluster pulsar population \citep{Camilo2005,Ransom2008}. For example, the globular cluster of the MW bulge, Terzan 5 is known to host at least 37 MSPs \citep{Terzan5}, whilst M28 has $\sim$ 8 MSPs out of a total PSR population of 12. \newline 

\noindent For the purposes of this work investigating the gravitational waveforms from astrophysical MSP-EMRBs, we take the 47 Tucane (`47 Tuc') as our example globular cluster. Indeed, 47 Tuc is known to host a substantial pulsar population \citep[25 pulsars, all of which have spin periods less than 8ms,][]{Pan2016} and it has been suggested -based on dynamical pulsar signatures- that at the core of 47 Tuc there also exists an intermediate mass black hole of mass $\sim 2.2 \times 10^3 M_{\odot}$ \citep{K2017NAT, TUC2017} although this interpretation has been disputed \citep{Mann2019}. The existence or otherwise of a IMBH in the centre of 47-Tuc is not an issue here - we just take 47 Tuc as a representative example of the sorts of globular clusters that could host MSP-EMRB. \newline 

\noindent Since the expected mass of the central BH is `intermediate' ($\sim 10^3 M_{\odot}$), in order to probe the gravitational strong field, the orbital periods of MSP in globular clusters systems are required to be shorter than those for the Galactic centre. An MSP in 47 Tuc with a 0.01 day orbital period and an eccentricity $e=0.9$ around a central IMBH of mass $\sim 2.2 \times 10^3 M_{\odot}$ would probe a gravitational potential of strength $\epsilon \sim 0.02$ as it passes through periapsis. Whilst no MSP-IMRB with an appropriately short orbital period has yet been detected, there remain real questions about whether our current PSR search algorithms are sophisticated enough to detect these weak signals from such relativistic systems in strong-field environments, when the signal is subject to a slew of general relativistic effects \citep{Psaltis2016,Kimpson2019b}. Similar to MSP-EMRB in the Galactic centre, the lack of MSP-IMRB detections in globular clusters is then likely due to inappropriate and insufficient observational methods. Going forward we set the BH mass as $\sim 2.2 \times 10^3 M_{\odot}$ and leave the spin parameter unchanged from the Galactic centre case, $a=0.6$. An advantage of 47 Tuc is that it is closer than the Galactic centre and so we set the observer to be at a distance of $4.0$ kpc. \newline

\begin{figure*}
	\includegraphics[width=\textwidth]{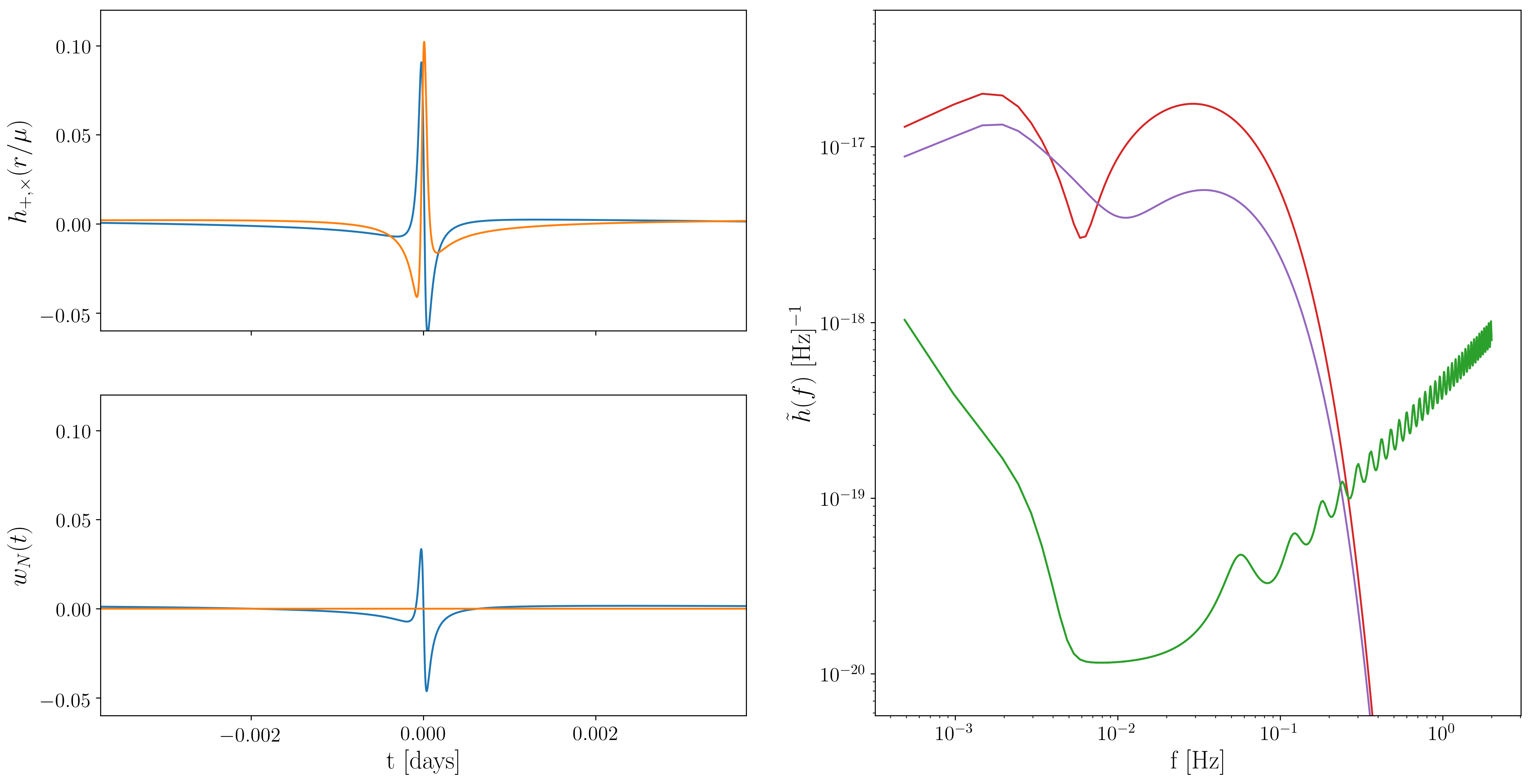}
	\caption{Waveforms in the time \textit{(left-side)} and frequency  \textit{(right-side)} domain for a MSP-EMRB going through periapsis in 47 Tuc. The orbital period is $P=0.01$ days with $e=0.9$, $\iota = 0 \deg$ and $\Omega = \pi/2$. The BH parameters are $M_{\rm BH} = 2.2 \times 10^3 M_{\odot}$, $a=0.6$. The observer is at $r_{\rm obs} = 4$ kpc,  $\Phi = 0$, with latitude $\Theta = 0$, (top left time waveform, red line frequency waveform) and $\Theta = \pi/2$ (bottom left time waveform, purple line frequency waveform). again for the time waveforms the blue and orange lines denote the $h_+, h_{\times}$ GW polarisations respectively. In the frequency spectra the green line is the LISA noise curve $\sqrt{S_n(f)}$ (not the characteristic strain $h_c$, Eq. \ref{eq:strain}) as described by Eq. \ref{eq:noise}.} \label{fig:Tuc}
\end{figure*}

\begin{figure}
	\includegraphics[width=\linewidth]{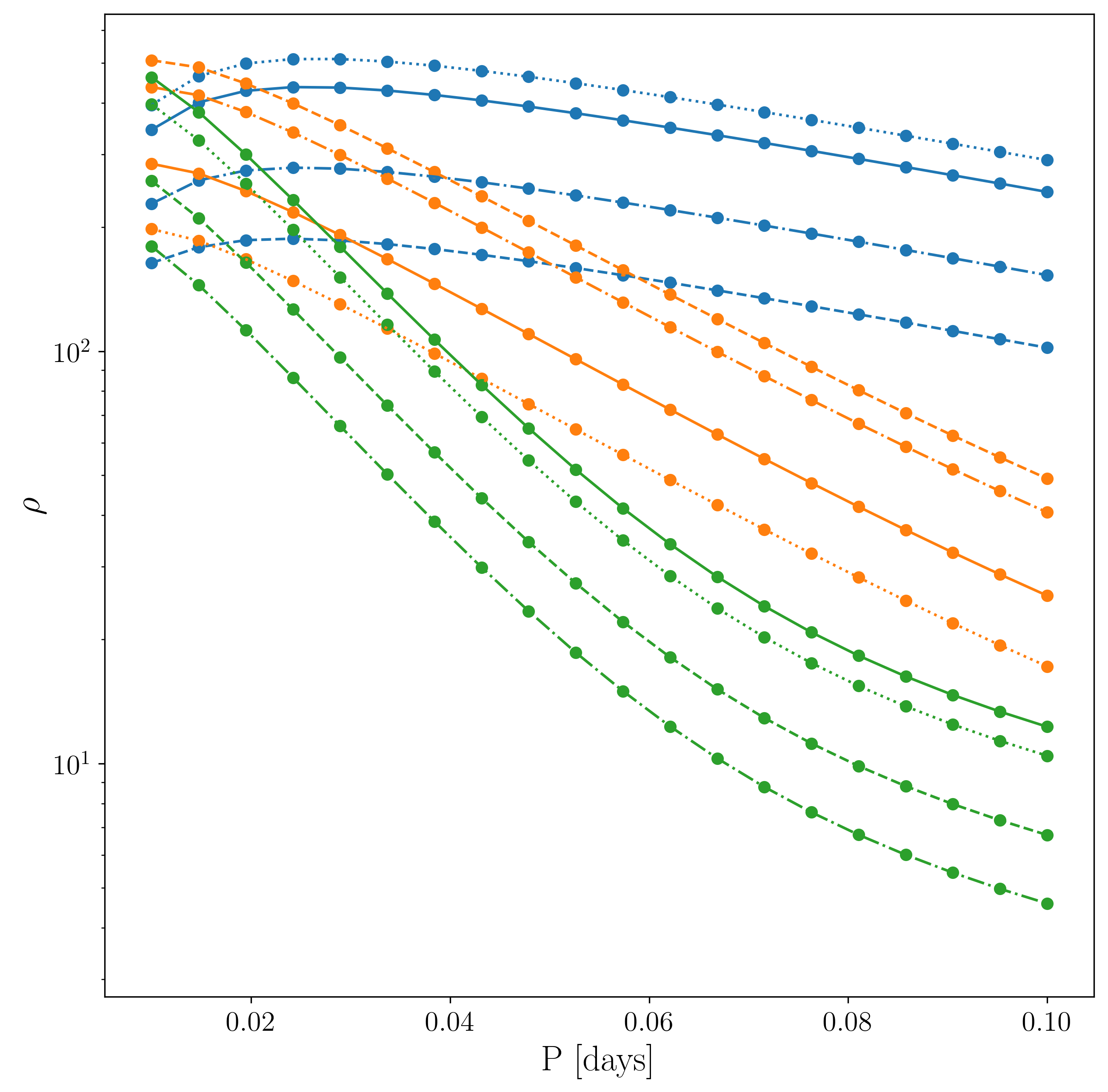}
	\caption{The signal to noise ratio of MSP-BH systems in 47-Tuc at $e=0.9,0.8,0.7$ (blue, orange,green respectively) at inclinations $\iota = 0,30,60,80 \deg$. Typically, shorter period orbits have greater SNRs, though this trend is less strong for more eccentric systems. The influence of the system orientation with respect to the observer is much stronger than for the Galactic centre systems, with face on observations corresponding to greater signal strengths.} \label{fig:SNRparamTuc}
\end{figure}

\noindent The time and frequency spectra of a MSP on an $P = 0.01$ day, $e=0.9$) orbit in this 47-Tuc system is shown in Fig. \ref{fig:Tuc}. Since the orientation of the BH spin axis with respect to the observer is unknown astrophysically, we  consider the two limiting cases; one where the observer location is the in the equatorial plane with respect to  the BH spin axis,  i.e. $\Theta = \pi/2$ and the other where the observer lies along the BH  axis such that $\Theta = 0$. We set the MSP to orbit in the equatorial plane ($\iota = 0$). 
In the $\Theta = 0$ case, $\rho = 396$, whilst when $\Theta = \pi/2$  the SNR is reduced to  $\rho= 144$. These SNRs are markedly higher than those calculated for the Galactic centre, even before exploring the full parameter space. This is on account of the fact that the systems are much closer and the signal occupies a different frequency regime on account of the different central BH masses and orbital periods. Whilst MSPs on sufficiently short orbits in globular clusters have not yet been discovered, less the existence of IMBH, this suggests that if thee systems do exist they offer fertile grounds for multimessenger strong field astronomy, without the complications of radio scattering and with increased SNRs due to their proximity and waveform frequencies. We can explore the parameter space analogous to how we did in the Galactic centre case. The results are presented in Fig. \ref{fig:SNRparamTuc}.
We can see that the SNR is strongest for more eccentric systems; all systems with $e=0.9$ and $P < 0.1$  days have $\rho > 100$. For these high eccentricities, $\rho$ drops as $P$ becomes very short due to the shift of the signal to a different part of the frequency spectrum where LISA is less sensitive. The SNR depends much more strongly on the system orientation than for Galactic centre systems, with greater values of $\rho$ for face-on systems.  \newline

\noindent Given that we have considered a nearby system, it is also interesting to investigate the detection prospects for a more distant, extragalactic system. Dwarf spheroidal galaxies are old clusters and also have exceptionally high densities in their central regions. For example M32, a satellite of Andromeda, has an exceptionally high measured stellar density in the central parsec of $3 \times 10^7 M_{\odot} \text{pc}^{-3}$ \citep{Lauer1992}. M32 is also thought to harbour a massive BH at its core of mass $\sim 3 \times 10^6 M_{\odot}$ \citep{Marel1997,Bender1996,Verolme2002,Bosch2010}. This combination of a massive BH in an old, dense stellar environment suggests that MSP-EMRBs should also be present in these systems. If the radio emission could be detected then these systems could also potentially be used for strong-field GR tests via radio timing. Whilst additional difficulties arise due to the distance of these sources (M32 is at a distance of $763$ kpc) and the commensurate faintness of the radio signal, there are compensatory advantages due to an expected lack of scattering. Again, target searches looking specifically for these systems with specialist search techniques for relativistic systems may prove fruitful. Indeed, whilst searches of nearby M32 did not definitely detect any pulsars, several single pulse events were detected which could be attributed to pulsar emission \citep{Rubio2013}. \newline 

\noindent For this work we take as M32 as our archetypal dwarf galaxy. Similar to when we were considering globular clusters and 47 Tuc, the orientation of the BH spin axis with respect to the observer is uncertain and so we again consider the two extremal cases where $\Theta = 0, \pi/2$. The BH mass is set as $3 \times 10^6 M_{\odot}$ and the observer at a distance of $763$ kpc. For an orbital period of $P = 0.01$ years,  $e=0.9$,  and a  favorable system orientation  ($\iota = 0$, $\Theta = 0$) the SNR is only $\rho = 0.3$. Naturally, those systems which are less favorably aligned have even lower SNRs. This low SNR of the GW is due to the same reason that radio pulsars are difficult to detect for these systems; they are simply very distant. In order to detect burst gravitational radiation would require orbital periods shorter than what we have considered here. For instance, if we  take the same system but shorten the orbital period to $ P = 0.001$ years $\rho = 14$. However such systems pass through periapsis at only $\sim 5 r_{\rm g}$ and their existence is unlikely, although not impossible. Alternatively, since these are regions of high stellar density it is possible that exotic systems could forms such as a  neutron star binary orbiting the massive central black hole i.e a extreme mass ratio hierarchical triple system \citep[e.g. ][]{Remmen2013}. The increased mass of the effective orbiter (i.e. 2 vs 1 neutron stars) would increase the resulting GW signal. Whilst the dynamics and resulting waveforms would be much more complicated in this triple case, a first order approximation to the SNR can be obtained since the NS separation is much smaller than the orbital radius and so we can approximate the the double NS system as equivalent to a single object of mass $2 \times 1.4 = 2.8 M_{\odot}$. Such a system again with $P =0.001$ would have an SNR of $\rho \sim 30$. Again, we grant that the formation and detection of such a system may be unlikely, however it is not completely unphysical and the huge scientific returns that could be returned from a multimessenger observations of a triple system in the  gravitational strong field make the possibility of detecting such a system at least worth some consideration.

\section{Discussion and Conclusions}
\label{sec:disco}
Through this work we have shown that in addition to MSP-EMRBs being used as strong-field GR probes through radio timing, with LISA it will be possible to detect bursting gravitational radiation from these systems as the MSP passes through periapsis. It is important to note that the gravitational waveforms have not been calculated in a self-consistent way - we have used a Numerical Kludge approach rather than a fully perturbative treatment. However, we are primarily concerned here with showing that GW bursts from typical MSP-EMRBs are detectable rather than calculating explicitly consistent waveforms. In addition, the NK approach is known to give strong agreement in the parameter space we have considered, and indeed we observe a high degree of overlap ($=1$) between our NK waveforms and those for which exact analytical solutions exist. We are therefore confident that the SNRs calculated in this work are at least reasonable approximations to the true burst SNRs. \newline

\noindent In order to use MSPs as strong-field probes, it is important to have as `clean' and environment as possible (c.f. complications from hydrodynamic drag, \citealt{Psaltis2012} or Newtonian perturbers, \citealt{Merritt2010}). Whilst the burst gravitational radiation is scientifically useful in and of itself, it also has the danger of acting as a potential noise source from the perspective of MSP radio timing. The gravitational radiation may influence the timing signal via two main channels. The first is via the effective perturbation that the gravitational wave introduces to the background spacetime metric which may in turn affect the geodesic of the photon ray emitted by the MSP. The second is by the GW emission influencing the MSP orbital motion. Both these points can be quickly addressed: GWs are transverse and so even for EM and GW radiation emitted coincidentally at periapsis, the photon ray will not be affected by the gravitational burst radiation. For the second point, due to the extreme mass ratio and the orbital periods considered for this work the orbital constants (e.g. $E,L$) do not meaningfully evolve over the burst duration and so will not impact either the PSR radio timing or - indeed - the calculated waveform SNRs. The lack of influence of the gravitational burst radiation on either the photon path or the orbital dynamics is advantageous from the perspective of radio pulsar timing and multimessenger astronomy since it means that the two messenger signals are entirely separate and will not influence each other. Therefore when calculating a PSR timing model the influence of this gravitational radiation will have no influence on the timing residuals. \newline

\noindent From these results, the most attractive target in terms of multimessenger strong-field astronomy is perhaps not the Galactic centre, but instead the centre of nearby globular clusters, such as 47 Tuc. MSP-EMRBs in globular systems would have shorter orbital periods and are also typically less distant in comparison to e.g. the Galactic centre. As a consequence the SNRs for the systems considered in this work were typically highest for globular cluster type systems, uncertainties in the system orientation (c.f BH spin axis, orbital inclination etc.) nonwithstanding. In addition to their higher GW burst SNRs, MSP-EMRB in globular clusters are also appealing from the perspective of MSP radio timing; globular clusters are known to host large populations of MSPs whilst radio observations do not suffer from line of sight effects due to astrophysical plasma causing scattering \citep[and temporal broadening of the pulse profile, e.g. ][]{Wucknitz2015} and spatial dispersion \citep{Kimpson2019a}. Moreover, since stars in globular clusters do not have a strong prograde/retrograde rotation preference, there is an even probability of detecting MSPs on a retrograde orbit, which would introduce additional interesting dynamical effects and imprints on the GW waveforms which we have not considered here. In addition to Globular clusters hosting sizable NS populations, there also exist numerous Globular clusters in the Galaxy which naturally increases the expected event rate. Globular clusters as MSP-EMRB targets do have their own drawbacks however, most notably that that the existence or otherwise of IMBH in the cores of these systems is far from well established. However, if a MSP on an appropriate orbit in a nearby globular cluster could be detected then the combination of simultaneous EM and GW multimessenger observations could firmly establish the existence - or otherwise - of IMBH, as well as probing other key questions of fundamental physics. \newline

\noindent In addition to globular clusters, observations of the Galactic centre remain a highly attractive target. In contrast to the centre of globular clusters, the existence of a massive nuclear BH associated with the Sgr A* radio source is well established \citep{Boehle2016,GravCollab2018}. The Galactic centre is also a region of high stellar density and is observed to host a collection of young massive OB stars, indicating a high rate of star formation. This observational evidence in conjunction with theoretical considerations of the historical star formation rate and the Galactic initial mass function suggests that the Galactic centre should host a large population of neutron stars \citep{Wharton2012}.
However despite numerous searches of the Galactic centre \citep[e.g.][]{Bates2011} no radio pulsars have been detected. Originally this dearth of detections was explained as being due to scattering due to astrophysical plasma along the line of sight causing pulse temporal smearing. Since PSR typically have steep radio spectra, searches are typically carried out at low radio frequencies. Unfortunately, at lower frequencies the scattering becomes more pronounced, hindering detections at the usual frequencies. However, even deep searches at higher frequencies \citep[e.g.][]{Macquart2010} returned no detections. Moreover, the detection in 2013 of the magnetar SGR J1745-29 in the inner parsec of the Galactic centre \citep{Kennea2013,Mori2013} indicates that the scattering towards the Galactic centre is less than expected and that previous searches should have detected pulsars in the Galactic centre. Such a tension has been dubbed the 'missing PSR problem' \citep{Dexter2014ApJ}. However, it was subsequently demonstrated \cite{Macquart2015,Rajwade2017} that if the Galactic centre pulsar population is composed primarily of MSPs then previous surveys were insensitive to the signals from this population. Moreover, if these MSP-EMRB inhabit the strong gravitational field regime, then their dynamics ans signals will be subject to strong field relativistic effects which current search methods may be not sufficiently sophisticated to detect. The hypothesis of a MSP GC population is not proposed merely as a solution to the missing PSR problem; there exists independent reasons to think that MSPs exist in this region. Firstly, as noted, the Galactic centre is a region of high stellar density. Consequently any pulsars would be subject to multiple close interactions, spinning up their rotation rate. Indeed, in the Globular cluster 47 Tuc - which is also a region of high density - $90 \%$ of the pulsars have spin periods less than $30$ ms. Secondly, the Galactic centre is known to have an overabundance of X-ray transients \citep{Hailey2018} which appear to be LMXB. As in globular clusters, the presence of a LMXB population is typically correlated with increased MSP numbers, given LMXB as a mechanism for the formation and spin up of MSPs. Finally, the well known Fermi $\gamma$-ray excess is best explained via a population of MSPs \citep{Yuan2014,Bartels2016}, rather than a signal due to dark matter annihilation. If MSP-EMRB in the Galactic centre do exist and can be detected and used for strong field tests of GR \citep[e.g.][]{Liu2012,Psaltis2016}, then this work illustrates that GW burst radiation should also be detectable, allowing another opportunity for strong field multimessenger astronomy. \newline 

\noindent As opposed to the two Galactic cases we have considered, the detection of GW burst radiation from extragalactic sources such as M32 is rather unlikely. This is simply on account of the distance to such sources reducing both the MSP radio signal and the GW signal. The SNR only becomes appreciable for more unlikely orbital configurations, for example short orbital periods in conjunction with some complex hierarchical triple system. Whilst such an instance may be more unlikely, it is not completely unphysical since the increased stellar density and associated two body interactions might allow for the formation of such exotic systems. However we do suggest that observational efforts are best concentrated on Galactic sources. \newline 

\noindent Beyond detecting these bursts, it is also of scientific interest to use them for parameter estimation of the central BH. Via standard blind GW astronomy, a GW burst from a 10 $M_{\odot}$ object at the Galactic centre is expected to be scientifically informative for appropriately short ($\lessapprox 10 r_{\rm g}$) orbits, in the best case providing mass and spin estimations of the central BH to one part in $10^4$ \citep{Berry2013MNRAS}. Due to their increased distance, for bursts from extragalactic sources the periapsis approach needs to be correspondingly shorter in order for these signals to be scientifically useful. Given this constraint these signals are also expected to be relatively rarer than Galactic ones. In the best case  parameters estimation of the central BH for extragalactic sources (e.g. M32) is expected to be at the $0.1 - 1 \%$ level \citep{Berry2013extragal}. \newline 

\noindent  These quoted uncertainties in the inferred parameters are derived with an uninformative prior, since in typical GW astronomy for burst sources there are few constrains on the source parameters. For a set of inferred parameters $\bar{\boldsymbol{\theta}}$, the associated uncertainty, in the limit of high SNR, is given by,
\begin{eqnarray}
\sqrt{\langle (\Delta \theta^i)^2\rangle} = \sqrt{\Sigma^{ii}}
\end{eqnarray} 
where for an uninformed prior $\bar{\boldsymbol{\Sigma}}$ is the inverse of the Fisher information matrix, $\bar{\boldsymbol{\Gamma}}$ \citep[see e.g.][]{Cutler1994}. However, in the presence of an accompanying, coincident radio signal we would have good estimates via a timing solution for some of the system parameters. In the case where the parameter has some Gaussian prior with a variance-covariance matrix $\boldsymbol{\bar{\Sigma}_0}$ then the variance-covariance of the posterior is 
\begin{eqnarray}
\bar{\boldsymbol{\Sigma}} = (\bar{\boldsymbol{\Gamma}} + \boldsymbol{\bar{\Sigma}_0}^{-1})^{-1}
\end{eqnarray}
and so apriori information will immediately improve the parameter estimation precision. Moreover, as noted in \cite{Cutler1994}, whilst apriori information can have a significant effect on the parameters to which the priors apply, it also aids the other parameters due to correlations. Electromagnetic pulsar timing observations would give constraining priors on a range of system parameters. Firstly, the BH mass and spin can be estimated via a pulsar timing solution. In the best case, over long timescales, these precisions are expected to be of the order $10^{-5} - 10^{-3}$ or better \citep{Liu2012}. Naturally even weaker constrains would be useful as a Bayesian prior. In addition estimates of the  quadrupole moment of the central BH can also be made \citep{Psaltis2016}. For pulsars - which are neutron stars - we immediately have a constrained estimate of the mass which is further improved as the timing solution is refined. Given the mass it is then possible to break the mass-distance degeneracy for bursts, and in any case the source distance will also be constrained independently through the host environment of the pulsar (e.g Galactic centre, 47 Tuc etc.). If the pulsar is observed for a sufficiently long time before the GW burst  we would also have good estimates of the coordinates of the orbital trajectory and quantities such as the orbital energy, angular momentum and Carter constant. Quantitative estimates of the improvement in the parameter uncertainty by using priors determined by radio pulsar timing observations require more involved calculations which are beyond the scope of this paper and we defer them for a future work. Furthermore, whilst the preceding discussion was focused on a single GW burst event, for bound orbiting systems multiple bursts would be observed on a timescale set by the orbital period, which would likely further improve the attainable measurement precision. \newline

\noindent An MSP-EMRB would be a unique gravitational wave source, given the presence of a continuous electromagnetic counterpart. Typical GW detection involves comparing a large number (covering the parameter space) of templates of the theoretical signal with the real noisy data. Such an approach would be unable to detect burst radiation from compact objects which are electromagnetically dark (e.g. BH, WD) over the sorts of orbital periods and timescales considered in this work. The presence of an EM beacon in the PSR beam is therefore a powerful tool for detecting gravitational radiation from these systems. Approaching the system from the other side i.e. the from the perspective of EM radio timing rather than GW astronomy, the potential to detect an accompanying gravitational radiation signal in conjunction with the radio MSP timing signal would enable true multimessenger astronomy in these strong-field regimes. The extent to which such observations could be used as a scientific apparatus c.f. parameter estimation from multimessenger observations has not been explored in this work but would be an interesting further development. For instance, given the detection of a MSP-EMRB and the associated EM and GW emission, to what precision can e.g. the BH spin be determined? We defer the investigation on the use of MSP-EMRB EM + GW burst waveforms for parameter estimation for a future work, but simply note here that the simultaneous electromagnetic and gravitational signals would provide a  unique astronomical probe of strong-field black hole spacetimes. It is also worth noting here that we have not explored the full orbital parameter space and the influence on the SNR, but instead just considered 3 major representative cases and briefly inspected the influence of system orientation. A full exploration of the parameter space we again defer for a future work. \newline 

\noindent For the typical orbital periods considered in this work $ P \sim 0.1$ year we have shown that radiation reaction self force effects can be neglected. As discussed this is advantageous from the perspective of MSP radio timing as a strong-GR probe since it means that the system remains `clean' and the gravitational radiation does not act as a noise source in the MSP timing residuals. However for shorter orbital periods with smaller periapsis distances the self-force backreaction will start to influence the orbital dynamics and burst waveform. Consistent modelling of Extreme Mass Ratio Inspirals (EMRIs) accounting for self force effects to second order is currently a major theoretical challenge \citep[see discussion in][]{Barack2019}. The resolution of this problem is key if EMRIs are to be detected with LISA and realise their scientific potential as precision strong-field probes.

%%%%%%%%%%%%%%%%%%%%%%%%%%%%%%%%%%%%%%%%%%%%%%%%%%



%\bibliographystyle{mnras}

\bibliographystyle{mnras}
\interlinepenalty=10000
\bibliography{bibfile}

%%%%%%%%%%%%%%%%%%%%%%%%%%%%%%%%%%%%%%%%%%%%%%%%%%

%%%%%%%%%%%%%%%%% APPENDICES %%%%%%%%%%%%%%%%%%%%%

%%%%%%%%%%%%%%%%%%%%%%%%%%%%%%%%%%%%%%%%%%%%%%%%%%

% Don't change these lines
\bsp	% typesetting comment
\label{lastpage}
\end{document}